\documentclass[prd,aps]{revtex4}
\usepackage{amssymb}

\newcommand{\re}{\mbox{Re}}

\newcommand{\Real}{\mathbb{R}}
\newcommand{\Complex}{\mathbb{C}}
\newcommand{\qed}{\hfill \fbox{} \vspace{.3cm}}
\newcommand{\sprod}[2]{\langle #1\, , \,#2 \rangle}
\newcommand{\dvrg}{{\bf div}}
\newcommand{\bbnab}{{\bf \nabla}}
\newcommand{\bbn}{{\bf n}}
\newcommand{\bbg}{{\bf g}}
\newcommand{\bbx}{{\bf x}}
\newcommand{\bby}{{\bf y}}
\newcommand{\bbA}{{\bf A}}
\newcommand{\bbB}{{\bf B}}
\newcommand{\bbE}{{\bf E}}
\newcommand{\bbF}{{\bf F}}
\newcommand{\bbJ}{{\bf J}}
\newcommand{\bbS}{{\bf S}}

\newcommand{\bbV}{{\bf V}}
\newcommand{\bbW}{{\bf W}}

\newtheorem{definition}{Definition}
\newtheorem{lemma}{Lemma}
\newtheorem{proposition}{Proposition}
\newtheorem{corollary}{Corollary}
\newtheorem{theorem}{Theorem}

\begin{document}

\title{A model problem for the initial-boundary value formulation of
Einstein's field equations}

\author{Oscar Reula}
\affiliation{Facultad de Matem\'atica, Astronom\'{\i}a y F\'{\i}sica,
Universidad Nacional de C\a'ordoba, Ciudad Universitaria,
Medina Allende y Haya de la Torre, C\'ordoba 5000, Argentina}

\author{Olivier Sarbach}
\affiliation{Department of Mathematics and Department of Physics and
Astronomy, Louisiana State University, 202 Nicholson Hall, Baton
Rouge, Louisiana 70803-4001, USA}
\affiliation{Theoretical Astrophysics 130-33,
California Institute of Technology, Pasadena, CA 91125, USA}

\begin{abstract}

In many numerical implementations of the Cauchy formulation of
Einstein's field equations one encounters artificial boundaries which
raises the issue of specifying boundary conditions. Such conditions
have to be chosen carefully. In particular, they should be compatible
with the constraints, yield a well posed initial-boundary value
formulation and incorporate some physically desirable properties like,
for instance, minimizing reflections of gravitational radiation.

Motivated by the problem in General Relativity, we analyze a model
problem, consisting of a formulation of Maxwell's equations on a
spatially compact region of spacetime with timelike boundaries. The
form in which the equations are written is such that their structure
is very similar to the Einstein-Christoffel symmetric hyperbolic
formulations of Einstein's field equations. For this model problem, we
specify a family of Sommerfeld-type constraint-preserving boundary
conditions and show that the resulting initial-boundary value
formulations are well posed. We expect that these results can be
generalized to the Einstein-Christoffel formulations of General
Relativity, at least in the case of linearizations about a stationary
background.
\end{abstract}

\maketitle

\section{Introduction}
\label{sect:intro}

For a number of years there has been an effort in the field of General
Relativity to obtain a set of boundary conditions which are consistent
with constraint propagation, physically reasonable and stable in the
sense that they yield a well posed initial-boundary value
formulation. Such a set is essential when integrating Einstein's field
equations on a domain with artificial timelike boundaries on a time
scale of astrophysical relevance, as is the case in most numerical
simulations of the binary black hole problem. Besides the success of
\cite{FN} for the case of their frame formulation, and in spite of a
lot of efforts \cite{CPBC-Stu, CPBC-CLT, CPBC-SSW, CPBC-SW, CPRST, CS,
CPBC-Frittelli, GMG1, GMG2} no boundary conditions which satisfy all
the above properties have been found for the more commonly used
tensorial formulations. In an attempt to understand this problem we
consider in section \ref{sect:MP} a toy model which in many aspects
resembles some of the evolution systems currently employed for
numerical evolution in General Relativity. This model is just
Maxwell's equations in a $3+1$ decomposition, but written as a first
order system in terms of the vector potential, the electric field, and
all first order spatial derivatives of the vector potential. The
allowance of all first order spatial derivatives of the vector
potential as variables, and not just the antisymmetric,
gauge-independent ones which describe the magnetic field, makes the
situation similar to the one on the above mentioned systems, where all
derivatives of the metric are indiscriminately promoted to evolution
variables, regardless of their gauge dependence. Thus, the Maxwell
system, which in gauge-independent variables (electric, and magnetic
fields) has a nice symmetric hyperbolic initial value formulation as a
set of evolution equations, here acquires most of the pathologies one
encounters for the general relativistic systems. In particular,
unphysical constraints appear, and the constraints propagate with
nontrivial speeds at timelike boundaries. The indeterminacy of the
evolution system under the addition of linear combination of
constraints is used to construct a two-parameter family of evolution
systems. For some values of the parameters the system is symmetric
hyperbolic, for some others it is only strongly hyperbolic, while
still for others it is neither, and so ill posed.  This system has
been used by a number of researchers in the past, in particular see
\cite{LS-FatMax}.

The construction of boundary conditions for this toy model is easier
than for the case of Einstein's equations, and we can look at it in
full generality. In particular, in the toy model, we know without
ambiguities what the physically relevant boundary conditions are which
lead to non-incoming radiation conditions, and so we can impose them,
together with some conditions ensuring constraint propagation. As
shown in section \ref{sect:CPBC} for a range of parameters this set of
boundary conditions satisfies the so called Kreiss condition
\cite{Kreiss, KL-Book, GKO-Book} and so they yield a well posed
problem for trivial initial data. But it turns out that, surprisingly,
this set of conditions does not yield well posed problems in the
traditional sense, that is when non-trivial initial data is
allowed. This is shown by a counterexample consisting of a solution to
the equations which does not satisfy the expected energy
estimates. This counterexample also indicates the source of the
problem which is, basically, the existence of static solutions which
are represented by linearly in time growing gauge potentials, and
motivates a new gauge condition.  By using this new condition, we are
able to show well posedness in a Hilbert space that controls the $L^2$
norm of the main variables and the constraint variables. In section
\ref{sect:EE} we derive a priori estimates. The main idea is to start
with an estimate for the gauge-invariant quantities instead of
estimating the norm given by the symmetrizer of the evolution system,
which is gauge dependent. Based on these estimates we show existence
of solutions in section \ref{sect:existence}. Since the new gauge
condition is imposed by some elliptic equations, some elliptic theory
is needed. For the proof of existence we employ the abstract theory of
semigroups, for which the elliptic aspects of the problem are taken
care of in a natural way. In section \ref{sect:Conclusions} we
summarize our results and discuss some of their implications for the
construction of boundary conditions for tensorial formulations of
Einstein's field equations. We have also provided an appendix with the
details of the elliptic theory needed; the material there is
standard. The main results and the counterexample that motivates the
new gauge condition are stated in the next section.

\section{Model problem and main results}
\label{sect:MP}

In this section we present our model problem which consists of a
system of evolution equations with constraints. We start with the case
without boundaries and state in which sense the resulting
initial-value problem is well posed. In particular, we introduce a
Hilbert space ${\cal H}$ which controls the $L^2$ norm of the main
{\em and} constraint variables. We then consider the presence of
artificial boundaries, discuss boundary conditions and show by an
explicit counterexample that the resulting initial-boundary value
problem is not well posed in the expected space ${\cal H}$. This
motivates a new gauge choice. Finally, we state the main result of
this article which proves well posedness in ${\cal H}$ of the
initial-boundary value problem for this new gauge choice.

Let $\Omega = \Real^3$, and denote by $\nabla$ the covariant
derivative with respect to the Eulerian metric ${\bf h} = \delta_{ij}
dx^i dx^j$ on $\Real^3$\footnote{Throughout this article, we use the
Einstein summation convention and raise and lower indices by means of
the metric ${\bf h}$.}. We are interested in the following evolution
system on $\Omega$:
\begin{eqnarray}
\partial_t \rho &=& \nabla^i J_i\; ,
\label{Eq:rhof}\\
\partial_t A_i &=& E_i + \nabla_i \phi, 
\label{Eq:Af}\\ 
\partial_t E_j &=& \nabla^i W_{ij} - (1+\alpha) \nabla^i W_{ji} + \alpha \nabla_j W
 + J_j\; ,
\label{Eq:Ef}\\ 
\partial_t W_{ij} &=& \nabla_i E_j + \frac{\beta}{2}\, h_{ij} \nabla^k E_k 
 + \nabla_i\nabla_j\phi - \frac{\beta}{2}\, h_{ij} \rho\; ,
\label{Eq:Wf}
\end{eqnarray}
where $\rho$ and $\phi$ are scalars on $\Omega$, $A_i$, $E_i$ and
$J_i$ are one-forms on $\Omega$, and $W_{ij}$ a two-tensor on $\Omega$
with trace $W = h^{ij} W_{ij}$. $\alpha$ and $\beta$ are two
parameters which determine the dynamics off the constraint
hypersurface which is defined by
\begin{eqnarray}
C &\equiv& \rho - \nabla^k E_k = 0,
\label{Eq:Gauss}\\
C^{(W)}_{ij} &\equiv& W_{ij} - \nabla_i A_j = 0.
\label{Eq:Cij}
\end{eqnarray}
Physically, the system
(\ref{Eq:Af},\ref{Eq:Ef},\ref{Eq:Wf},\ref{Eq:Gauss},\ref{Eq:Cij}) is
equivalent to Maxwell's equations on Minkowski space, where $\phi$
represents the electrostatic potential, $A_i$ the vector potential,
$E_i$ the electric field, $(B_i) =
(W_{23}-W_{32},W_{31}-W_{13},W_{12}-W_{21})$ the magnetic field, and
$\rho$ and $J_j$ represent the charge and current density,
respectively, which obey the continuity equation
(\ref{Eq:rhof}). Notice that for given current density,
Eq. (\ref{Eq:rhof}) can be integrated separately in order to obtain
$\rho$, which can then be used as a source function in order to
integrate the equations
(\ref{Eq:Af},\ref{Eq:Ef},\ref{Eq:Wf}). However, we will find it more
convenient to interpret $\rho$ as a field being evolved along with the
fields $A_i$, $E_i$, $W_{ij}$ since in this case the constraint
variable $C$ depends linearly on the evolution fields. We assume that
the electrostatic potential and the current density are a priori
given. The motivation for introducing the fields $W_{ij}$, which
represent all the first order spatial derivatives of the vector
potential, instead of using the magnetic field is to obtain a system
of equations whose structure is similar to the one of the
Einstein-Christoffel formulation of Einstein's field equations
\cite{FR, AY, Hern, KST}. In the latter, one has evolution equations
for the components of the three-metric $g_{ij}$, the extrinsic
curvature $K_{ij}$, and some symbols $d_{kij}$ that are linear
combinations of the Christoffel symbols. These fields are subject to
the Hamiltonian and momentum constraints, and to the constraints
$d_{kij} - \partial_k g_{ij} = 0$. The structure of the equations is
very similar to the ones in our model problem with the correspondence
$g_{ij} \leftrightarrow A_i$, $K_{ij} \leftrightarrow E_i$, $d_{kij}
\leftrightarrow W_{ij}$, $N^i \leftrightarrow \phi$, where $N^i$ is
the shift vector field.

In the following, we impose the condition $\alpha\cdot \beta > 0$
which is a necessary and sufficient condition for the evolution system
(\ref{Eq:rhof},\ref{Eq:Af},\ref{Eq:Ef},\ref{Eq:Wf}) to be strongly
hyperbolic which in turn yields well posedness of the associated
Cauchy problem in $L^2(\Omega)$ \cite{KL-Book}. The constraints'
propagation is described by the following evolution system
\begin{eqnarray}
\partial_t C^{(W)}_{ij} &=& -\frac{\beta}{2}\, h_{ij} C,
\label{Eq:Cijf}\\ 
\partial_t C &=& -\alpha \nabla^k C_k\; ,
\label{Eq:Cf}\\ 
\partial_t C_k &=& -\beta\nabla_k C,
\label{Eq:Ckf}
\end{eqnarray}
which is a consequence of the evolution system
(\ref{Eq:rhof},\ref{Eq:Af},\ref{Eq:Ef},\ref{Eq:Wf}). Here, we have
introduced the new constraint variable $C_k \equiv 2h^{ij}\nabla_{[k}
C^{(W)}_{i]j} = \nabla_k W - \nabla^j W_{kj}$ in order to obtain a
first order evolution system.  A simple energy estimate shows that the
unique solution to Eqs. (\ref{Eq:Cijf},\ref{Eq:Cf},\ref{Eq:Ckf}) with
trivial initial data is trivial, and therefore, the constraints remain
satisfied if satisfied initially. Summarizing, we have the following

\begin{theorem}[Well posedness of the Cauchy problem]
\label{Thm:Cauchy}
For $\alpha\beta > 0$ the constrained Cauchy problem associated with
(\ref{Eq:rhof},\ref{Eq:Af},\ref{Eq:Ef},\ref{Eq:Wf}),
(\ref{Eq:Gauss},\ref{Eq:Cij}) on $\Omega = \Real^3$ is well posed in
the following sense: Let $\tau > 0$ be an arbitrary fixed constant
with dimension of length, and write $u(t) = (\tau\rho(t),\tau^{-1}
A_j(t), E_i(t), W_{ij}(t))$ (main variables), $v(t) =
(\tau^{-1}C^{(W)}_{ij}(t),C(t),C_k(t))$ (constraint variables) and
$j(t) = (J_i(t),\tau\nabla^i
J_i(t),\tau^{-1}\nabla_i\phi,\nabla_i\nabla_j\phi)$ (source
functions). Given smooth source functions $j(t)\in L^2(\Omega)$, $t >
0$ such that $t \mapsto j(t)$ is continuous, and given smooth initial
data $u_0 \in L^2(\Omega)$, there exists a unique solution $u(t)\in
L^2(\Omega)$ of the evolution system
(\ref{Eq:rhof},\ref{Eq:Af},\ref{Eq:Ef},\ref{Eq:Wf}) with $u(0) =
u_0$. Furthermore, the solution obeys the estimate
\begin{equation}
\| u(t) \|_{L^2(\Omega)}^2 \leq a\, e^{b t/\tau} \left[ \| u_0 \|_{L^2(\Omega)}^2 + 
\tau \int_0^t \| j(s) \|_{L^2(\Omega)}^2 ds \right],
\label{Eq:L2CauchyEstimate}
\end{equation}
for some constants $a$, $b$.

If the initial data is such that $v_0\in L^2(\Omega)$, then $v(t)\in
L^2(\Omega)$ and we also have the estimate
\begin{equation}
\| v(t) \|_{L^2(\Omega)}^2 \leq c\, e^{d t/\tau} \| v_0 \|_{L^2(\Omega)}^2,
\end{equation}
for some constants $c$, $d$. In particular, this implies that initial
data which satisfies the constraints initially automatically satisfies
the constraints at later time.
\end{theorem}

{\bf Remarks}:
\begin{enumerate}
\item
The constant $\tau$ is an artificial length scale that is introduced
in order to avoid adding quantities which have different units. By
choosing this scale to be arbitrarily large we can make the growth
rate $1/\tau$ in the estimates as small as we like.
\item
The precise sense in which there exists a solution $u(t)\in
L^2(\Omega)$ is in the sense of a strongly continuous semigroup in
$L^2(\Omega)$ whose generator is determined by the operator on the
right-hand side of (\ref{Eq:rhof},\ref{Eq:Af},\ref{Eq:Ef},\ref{Eq:Wf}).
\item
In the following, we will replace the Hilbert space $\{ u =
(\tau\rho,\tau^{-1}A_i,E_i,W_{ij}) \in L^2(\Omega) \}$ by the Hilbert
space
\begin{equation}
{\cal H} = \{ u = (\tau\rho,\tau^{-1} A_i,E_i,W_{ij}) \in L^2(\Omega) : v = (\tau^{-1} C^{(W)}_{ij}, C, C_k) \in
L^2(\Omega) \}
\end{equation}
with scalar product $\sprod{u_1}{u_2}_{\cal H} = (u_1,
u_2)_{L^2(\Omega)} + \tau^2(v_1,v_2)_{L^2(\Omega)}$. This Hilbert
space controls the $L^2$ norm of the main variables {\em and} the
constraint variables. In this case, we replace the estimate
(\ref{Eq:L2CauchyEstimate}) by the estimate
\begin{equation}
\| u(t) \|_{\cal H}^2 \leq a\, e^{b t/\tau} \left[ \| u_0 \|_{\cal H}^2 + 
\tau\int_0^t \| j(s) \|_{L^2(\Omega)}^2 ds \right].
\label{Eq:HCauchyEstimate}
\end{equation}
The space ${\cal H}$ will be important when boundaries are present,
where we will be able to show well posedness in ${\cal H}$ (see
Theorem \ref{Thm:Main} below) but not in $L^2$. That is, we will be
able to derive an estimate which involves the $L^2$ norms of the main
{\em and} the constraint variables, but not the main variables alone.
\end{enumerate}

The purpose of this article is to analyze the constrained evolution
system (\ref{Eq:rhof},\ref{Eq:Af},\ref{Eq:Ef},\ref{Eq:Wf}),
(\ref{Eq:Gauss},\ref{Eq:Cij}) on a bounded domain $\Omega$ of
$\Real^3$ with $C^\infty$ boundary $\partial\Omega$. In the following,
$\bbn = n_i dx^i$ denotes the outward unit one-form to
$\partial\Omega$. We also introduce the projection operator $H_i^j =
\delta_i^j - n_i n^j$ on the tangent space of $\partial\Omega$. For
technical reasons, we assume that the extrinsic curvature of the
boundary surface, defined by $\kappa_{ij} = H_i^r H_j^s \nabla_r n_s$,
is positive semi-definite at each point of the boundary.

We wish to specify boundary conditions on $\partial\Omega$ which imply
a well posed initial-boundary value problem in ${\cal H}$. In
particular, these conditions have to be specified such that the
constraints are propagated which means that the constraints
(\ref{Eq:Gauss},\ref{Eq:Cij}) should hold at later times if satisfied
initially. We shall call boundary conditions which have this property
{\em constraint preserving}. In order to specify such conditions we
impose homogeneous maximally dissipative boundary conditions for the
constraints' propagation system
(\ref{Eq:Cijf},\ref{Eq:Cf},\ref{Eq:Ckf}), which means that we couple
the in- to the outgoing characteristic fields with a smooth coupling
function $c$ on $\partial\Omega$ with the property that $|c|\leq 1$:
\begin{equation}
V^{(+)} = c\; V^{(-)},
\label{Eq:CPBC}
\end{equation}
where
\begin{displaymath}
V^{(+)} \equiv C - \frac{\sqrt{\alpha\beta}}{\beta}\, n^k C_k\; ,\qquad
V^{(-)} \equiv C + \frac{\sqrt{\alpha\beta}}{\beta}\, n^k C_k\; ,
\end{displaymath}
are the in- and outgoing fields, respectively. A simple energy
argument (see the next section) shows that the unique solution to
Eqs. (\ref{Eq:Cijf},\ref{Eq:Cf},\ref{Eq:Ckf}) with the boundary
condition (\ref{Eq:CPBC}) and trivial initial data is
trivial. Therefore, the boundary conditions (\ref{Eq:CPBC}) are
constraint preserving. Furthermore, we impose the boundary conditions
\begin{equation}
w_i^{(+)} = S_i^{\;\; j} w_j^{(-)} + g_i\; , \qquad
w_i^{(\pm)} \equiv H_i^{\;\; j} E_j \pm n^k (W_{ki} - W_{ik}),
\label{Eq:PoyntingBC}
\end{equation}
where $S_i^{\;\; j}$ is a smooth matrix-valued function on
$\partial\Omega$ with the properties that $S_i^{\;\; j} n^i = 0$,
$S_i^{\;\; j} n_j = 0$ and $H^{ij} S_i^{\;\; r} S_j^{\;\; s} \leq
H^{rs}$ if $g_i = 0$ and $H^{ij} S_i^{\;\; r} S_j^{\;\; s} \leq \delta
H^{rs}$ for some $\delta < 1$ otherwise. The function $g_i$ is a
smooth function on $\partial\Omega$ with the property that $n^i g_i =
0$. Physically, the boundary condition (\ref{Eq:PoyntingBC}) permits
to control the normal component of the Poynting vector $P_n = -n^i E^j
(W_{ij} - W_{ji})$: When $g_i = 0$, this condition makes sure that
$P_n$ is nonnegative, meaning that the total energy flux through the
boundary is nonnegative. When $S_i^{\;\; j} = 0$ the data $g_j$
permits to introduce a wave which travels in normal direction towards
the boundary.

With these boundary conditions one would expect to have an estimate, 
and the corresponding theorem of uniqueness and existence, of the 
following form:
\begin{eqnarray}
&& \| u(t) \|_{\cal H}^2  \leq a\, e^{b t/\tau} \left[ \| u_0 \|_{\cal H}^2 
 + \tau\int_0^t \| j(s) \|_{L^2(\Omega)}^2 ds
 + \int_0^t \| \bbg(s) \|_{L^2(\partial\Omega)}^2 ds \right],
\nonumber\\
&& \| v(t) \|_{L^2(\Omega)}^2 \leq a\, e^{b t/\tau} \| v_0 \|_{L^2(\Omega)}^2,
\nonumber
\end{eqnarray}
for some constants $a$, $b$. But this expectation is false, as shows
the following counterexample.

\subsection{Ill posedness in $L^2$}

Suppose that the source functions $\phi$ and $J_j$ vanish identically,
and consider the following family of solutions:
\begin{eqnarray}
\rho &=& 0, \nonumber\\
A_i &=& t\nabla_i f, \nonumber\\
E_i &=& \nabla_i f, \label{Eq:SolNotL2}\\
W_{ij} &=& t\nabla_i\nabla_j f \nonumber, 
\end{eqnarray}
where $f$ is a smooth, time-independent, harmonic function
($\nabla^k\nabla_k f = 0$). This family satisfies the constraints
(\ref{Eq:Gauss},\ref{Eq:Cij}), the evolution equations
(\ref{Eq:rhof},\ref{Eq:Af},\ref{Eq:Ef},\ref{Eq:Wf}) and the boundary
condition (\ref{Eq:CPBC}). It has the initial data
\begin{displaymath}
\left. \rho \right|_{t=0} = 0, \qquad
\left. A_i \right|_{t=0} = 0, \qquad
\left. E_i \right|_{t=0} = \nabla_i f, \qquad
\left. W_{ij} \right|_{t=0} = 0,
\end{displaymath}
and satisfies the boundary conditions (\ref{Eq:PoyntingBC}) with
boundary data given by
\begin{displaymath}
g_i = w_i^{(+)} - S_i^{\;\; j} w_j^{(-)} 
 = \left. (H_i^{\;\; j} - S_i^{\;\; j}) \nabla_j f  \right|_{x=0} \; .
\end{displaymath}
This family of solutions does not obey the desired estimate in ${\cal
H}$ since $W_{ij}$ depends on second derivatives of $f$ whereas the
initial and boundary data depend only on first derivatives of $f$\footnote{In
order to see this more explicitly, assume that $\Omega = \Real_+
\times S^1 \times S^1$ where the $y$ and $z$ directions are periodic
with period $1$, and consider the family of solutions parametrized by
$$
f_n(x,y,z) = e^{-\omega_n(x + iy)}, 
$$
where $\omega_n = 2\pi n$ and $n$ is a fixed integer. Clearly, $f_n$
is harmonic. Furthermore,
$$
I \equiv \| E_i \|^2_{L^2(\Omega)} = 2\omega_n^2\int_0^\infty e^{-2\omega_n x} dx
 = \omega_n,
\qquad
B \equiv \| H_i^{\;\; j} \nabla_j f \|^2_{L^2({[0,t]}\times\partial\Omega)} 
 = \omega_n^2\int_0^t 1\; dt = t\omega_n^2.
$$
However, we have
$$
N \equiv \| W_{ij}(t) \|^2_{L^2(\Omega)} 
 = 4t^2\omega_n^4\int_0^\infty e^{-2\omega_n x} dx = 2t^2\omega_n^3.
$$ 
Therefore, for each fixed $t$, $N/I = 2t^2\omega_n^2 \rightarrow
\infty$ and $N/B = 2t\omega_n \rightarrow \infty$ as $n
\rightarrow\infty$ so the system is ill posed in ${\cal
H}$.}. Physically, the solution (\ref{Eq:SolNotL2}) represents an
electrostatic solution in a ``bad'' gauge corresponding to a boundary
with charge density proportional to $n^i\nabla_i f$. This gauge is
``bad'' in the sense that it is not adapted to electrostatic solutions
since it requires the electric potential to be zero. As a consequence,
the potential grows linearly in time. Notice that the solution
(\ref{Eq:SolNotL2}) is trivial in the absence of boundaries since (if
one requires the initial data to decay sufficiently fast at infinity)
$f$ has to be zero. This is compatible with the fact that in those
cases one can show well posedness in $L^2$ since the evolution system
is strongly hyperbolic (see Theorem \ref{Thm:Cauchy}).

\subsection{Main result}

The above counterexample shows that the simple gauge condition
$\phi=0$ does not lead to a problem that is well posed in ${\cal
H}$. This counterexample motivates the following condition on $\phi$:
\begin{equation}
\nabla^k\nabla_k\phi = -\nabla^k E_k\; ,\qquad
n^k\nabla_k\phi = -n^k E_k\quad \hbox{on $\partial\Omega$}.
\label{Eq:EllipticGauge}
\end{equation}
This condition precludes the type of solution leading to the
counterexample, and, as we will state shortly, allows for the
formulation of a well posed initial-boundary value problem. Notice
that the gauge condition (\ref{Eq:EllipticGauge}) guarantees that
$\partial_t (n^k A_k) = 0$ on $\partial\Omega$. In particular, this
implies that the solution satisfies $n^k A_k = 0$ on $\partial\Omega$
if the initial data satisfies this condition. In order to state the
main result of this article, we introduce the Hilbert space
\begin{equation}
{\cal H}' = \{ u\in {\cal H} : n^k A_k = 0 \hbox{ on $\partial\Omega$} \}.
\end{equation}

\begin{theorem}[Well posedness of the initial-boundary value problem]
\label{Thm:Main}
Consider the constrained evolution system
(\ref{Eq:rhof},\ref{Eq:Af},\ref{Eq:Ef},\ref{Eq:Wf}), (\ref{Eq:Gauss},\ref{Eq:Cij})
on a bounded domain $\Omega$ with $C^\infty$ boundary and boundary
conditions (\ref{Eq:CPBC},\ref{Eq:PoyntingBC}), where $\phi$ is
determined by the elliptic system (\ref{Eq:EllipticGauge}). Assume
that $\alpha\beta > 0$, and that the extrinsic curvature of the
boundary surface, $\kappa_{ij}$, is positive semi-definite at each
point of the boundary. Assume that the matrix-valued function 
$S_i^{\;\; j}$ appearing in the boundary condition (\ref{Eq:PoyntingBC})
has the form $S_i^{\;\; j} = s H_i^{\;\; j}$ for some function $s\in C^\infty(\partial\Omega)$,
and assume that $|c| < 1$, $|s| < 1$ on $\partial\Omega$.

Then, given smooth source functions $\hat{j}(t) = (J_i(t), \nabla^i
J_i(t))$, $t > 0$, in $L^2(\Omega)$, such that $t\mapsto\hat{j}(t)$ is
continuous, and given smooth initial data $u_0 \in {\cal H}'$ and
smooth boundary data $g_i(t)$ in $L^2(\partial\Omega)$ with
appropriate compatibility conditions (see
Eqs. (\ref{Eq:CC1},\ref{Eq:CC2}) below) at $\{ t=0 \} \times
\partial\Omega$, there exists a unique solution $u(t)\in {\cal H}'$ of
(\ref{Eq:rhof},\ref{Eq:Af},\ref{Eq:Ef},\ref{Eq:Wf}),
(\ref{Eq:EllipticGauge}), (\ref{Eq:CPBC},\ref{Eq:PoyntingBC}) with
$u(0)=u_0$. Furthermore, the solution obeys the estimates
\begin{eqnarray}
&& \| u(t) \|_{\cal H}^2  \leq a\, e^{b t/\tau} \left[ \| u_0 \|_{\cal H}^2 
 + \tau\int_0^t \| \hat{j}(s) \|_{L^2(\Omega)}^2 ds 
 + \int_0^t \| \bbg(s) \|_{L^2(\partial\Omega)}^2 ds \right],
\label{Eq:energy_estimate1}\\
&& \| v(t) \|_{L^2(\Omega)}^2 \leq a\, e^{b t/\tau} \| v_0 \|_{L^2(\Omega)}^2.
\label{Eq:energy_estimate2}
\end{eqnarray}
\end{theorem}

{\bf Remarks}:
\begin{enumerate}
\item
Again, the precise sense in which there exists a solution $u(t)\in
{\cal H}'$ is in the sense of the existence of a strongly continuous
semigroup in ${\cal H}'$. This is discussed in section
\ref{sect:existence}.
\item
The assumptions on the nonnegativeness of $\kappa_{ij}$ and on the
special form of the coupling matrix $S_i^{\;\; j}$ can be weakened, but
the proof is technically more complicated in those cases.
\end{enumerate}

In the next section, we start with some preliminary investigations of
the initial-boundary problem
(\ref{Eq:rhof},\ref{Eq:Af},\ref{Eq:Ef},\ref{Eq:Wf}),
(\ref{Eq:CPBC},\ref{Eq:PoyntingBC}) which are valid in the
``high-frequency limit''. That is, we consider the case where $\Omega$
is a half plane and derive some necessary conditions for well
posedness, using Laplace and Fourier transformation techniques. A
complete proof of Theorem \ref{Thm:Main} is given in the subsequent
sections.

\section{Constraint-preserving boundary conditions}
\label{sect:CPBC}

In this section we construct a family of constraint-preserving
boundary conditions for the case where $\Omega = \{ (x,y,z)\in\Real^3
: x > 0 \}$ is the half space. We apply Laplace-Fourier techniques in
order to derive necessary conditions for well posedness of the
resulting initial-boundary value problem. We also derive a stronger
condition, known as the Kreiss condition, which yields an $L^2$
estimate in the case of trivial initial data and trivial
sources. However, the counterexample presented in the previous section
shows that this result cannot be generalized to the case of
non-trivial initial data when the temporal gauge $\phi=0$ is
adopted. This shows that the verification of the Kreiss condition for
constraint-preserving boundary conditions does not necessarily yield
well posedness in the space $L^2$. For simplicity we only consider the
source-free case here, where $\rho=0$, $J_j=0$. We also neglect the
evolution equations for $A_i$ and $C^{(W)}_{ij}$ in this section since
they can be integrated separately once a solution of
(\ref{Eq:Ef},\ref{Eq:Wf}) and (\ref{Eq:Cf},\ref{Eq:Ckf}) has been
obtained.

In order to construct constraint-preserving boundary conditions, we
first find the characteristic speeds and fields with respect to the
unit outward normal $(n_i) = (-1,0,0)$. For the evolution system
(\ref{Eq:Ef},\ref{Eq:Wf}) these are
\begin{eqnarray}
& \pm r,  & v_x^{(\pm)} \equiv E_x \mp \frac{r}{\beta}\; W^A_{\;\; A}\\
& \pm 1,  & v_A^{(\pm)} \equiv E_A \mp (W_{xA} - (1+\alpha)W_{Ax}))\\
& 0,      & W_{Ax}\\
& 0,      & \chi_{AB} \equiv W_{AB} - \frac{1}{2}\,\delta_{AB} W^C_{\;\; C}\\
& 0,      & \kappa \equiv \beta\, W_{xx} - (1 + \beta/2) W^A_{\;\; A},
\end{eqnarray}
where here and in the following, Capital indices refer to the
directions $y$ and $z$ which are transversal to the boundary,
$\delta_{AB}=1$ if $A=B$ and zero otherwise, and $r =
\sqrt{\alpha\beta} > 0$. For the constraints' propagation system
(\ref{Eq:Cf},\ref{Eq:Ckf}), the characteristic speeds and fields are
\begin{eqnarray}
& \pm r,  & V^{(\pm)} \equiv C \pm \frac{r}{\beta}\; C_x\\
& 0,      & C_A\; .
\end{eqnarray}
Defining the energy norm
\begin{equation}
{\cal E}_c = \frac{1}{2}\int_{\Omega} 
\left( C^2 + \frac{\alpha}{\beta}\, C^k C_k \right) d^3 x,
\end{equation}
taking a time derivative, using Eqs. (\ref{Eq:Cf},\ref{Eq:Ckf})
and integration by parts, we find
\begin{equation}
\frac{d}{dt} {\cal E}_c = -\alpha\int_{\Omega} 
\left( C \nabla^k C_k + C^k \nabla_k C \right) d^3 x
 = \alpha\int_{x=0} C C_x\, dy dz 
 = \frac{r}{4} \int_{x=0} \left( (V^{(+)})^2 - (V^{(-)})^2 \right) dy dz.
\end{equation}
Therefore, if we impose the boundary condition
\begin{equation}
V^{(+)} = c\; V^{(-)}
\label{Eq:CPBCflat}
\end{equation}
with $|c|\leq 1$, it follows that the positive definite quantity
${\cal E}_c$ cannot increase; in particular a solution with initial
data such that $C=0$, $C_k = 0$ satisfies $C=0$ and $C_k=0$ at later
times as well. Re-expressing condition (\ref{Eq:CPBCflat}) in terms of
the variables of the main system yields
\begin{eqnarray}
0 &=& r(V^{(+)} - c\; V^{(-)})
\nonumber\\
  &=& \partial_t( v_x^{(+)} + c\; v_x^{(-)}) 
   + \nabla^A\left[ \frac{r}{\beta}(1-c) E_A + (1+c) F_{xA} - \frac{r}{\beta}(1-c)\nabla_A\phi \right],
\label{Eq:CPBCMainVars}
\end{eqnarray}
where $F_{xA} = W_{xA} - W_{Ax}$, and where we have used the main
evolution equations (\ref{Eq:Ef},\ref{Eq:Wf}) in order to trade normal
derivatives ($\partial_x$) by derivatives that are tangential to the
boundary ($\partial_t$, $\partial_A$). We can interpret this equation
as an evolution equation for $v_x^{(+)} + c\; v_x^{(-)}$ {\em at the
boundary}. Since the main evolution system has three ingoing modes, we
need two more boundary conditions. One possibility is to require a
maximally dissipative boundary condition on the transversal fields
\begin{equation}
v_A^{(+)} = S_A^{\;\; B}\; v_B^{(-)} + g_A\; ,
\end{equation}
where $g_y$, $g_z$ are a priori specified functions on the boundary and
where the $2\times 2$ matrix $S = (S_A^{\;\; B})$ satisfies $S^\dagger
S \leq 1$. From a physical point of view it is more convenient to
require
\begin{equation}
E_A - F_{xA} = S_A^B(E_B + F_{xB}) + g_A\; ,
\end{equation}
since this controls the radiation flux through the boundary, as
discussed in the previous section. In order to explore both
possibilities, we shall analyze the family of boundary conditions
consisting of Eq. (\ref{Eq:CPBCflat}) and
\begin{equation}
\left[ E_A - W_{xA} + (1+a)W_{Ax} \right]  = S_A^B\left[ E_B + W_{xB} - (1+a)W_{Bx} \right] + g_A\; ,
\label{Eq:Phys}
\end{equation}
where the parameter $a$ is equal to either $\alpha$ or zero.


\subsection{Particular cases}


Here, we assume that the parameters $\alpha$ and $\beta$ satisfy $-2 <
\alpha < 0$, $\beta < -2/3$ which makes sure that the evolution system
(\ref{Eq:Ef},\ref{Eq:Wf}) is symmetric hyperbolic. Suppose that $\phi$
satisfies $\nabla^A\nabla_A\phi = 0$ at the boundary. If we choose
$c=-1$ and $S_A^B = -\delta_A^B$, the boundary conditions reduce to
\begin{equation}
v_x^{(+)} = v_x^{(-)} + g,
\qquad
v_A^{(+)} = -v_A^{(-)} + g_A\; ,
\end{equation}
where, given $g_A$, the source function $g$ is obtained by integrating
the equation $\partial_t g = -r\beta^{-1}\nabla^A g_A$ at the
boundary. Once $g$ is determined the resulting initial-boundary value
problem is well posed since we specify maximally dissipative boundary
conditions for a symmetric hyperbolic system. Notice that $v_A^{(+)} +
v_A^{(-)} = 2E_A$, so if $g_A=0$ this means that the boundary is a
conductor.

If we choose, instead, $c=1$, $S_A^B = \delta_A^B$ and $a = \alpha$,
we obtain
\begin{equation}
v_x^{(+)} = -v_x^{(-)} + g,
\qquad
v_A^{(+)} = v_A^{(-)} + g_A\; ,
\end{equation}
where now $g$ is determined by the following evolution system at the
boundary (given $g_A$ and $\nabla_x\phi$ at the boundary):
\begin{eqnarray}
\partial_t g + 2\alpha\nabla^A W_{Ax} &=& \nabla^A g_A\, ,\\
\partial_t W_{Ax} - \frac{1}{2}\,\nabla_A g &=& \nabla_A\nabla_x\phi.
\end{eqnarray}
Since $-2 < \alpha < 0$, this boundary evolution system is well posed,
and it can be integrated separately to obtain the boundary function
$g$ (and the zero speed field $W_{Ax}$). The functions $g$ and $g_A$
are then used as boundary data to integrate the bulk system, and the
resulting initial-boundary value problem is well posed. However, in
this case, the physical interpretation is less clear since the free
data $g_A = -2F_{xA} + 2\alpha W_{Ax}$ is not gauge-invariant.

These two sets of well posed constraint-preserving boundary conditions
have been generalized to the linearized Einstein-Christoffel
formulation of Einstein's equations \cite{CPRST}. However, in order to
construct a radiative-type boundary condition, as described in the
previous section, we need to choose $S_A^B = 0$ and $a = 0$ in
Eq. (\ref{Eq:Phys}), and in this case the well posedness of the
resulting system is more involved since the resulting boundary
conditions are not in maximal dissipative form. We analyze this in the
next subsection.


\subsection{Fourier-Laplace analysis}


Since the equations and boundary conditions are linear and have
constant coefficients, we can solve the initial-boundary value problem
by performing a Laplace transformation in time and a Fourier
transformation in the spatial directions which are tangential to the
boundary. In other words, we are considering solutions of the form
$f(x) e^{st + i\omega_A x^A}$, where $s$ is a complex number with
positive real part, $(\omega_A) = (\omega_y,\omega_z)$ is a real
two-vector and $f\in L^2(\Real_+)$. For given boundary data there must
be a unique such solution for each $\re(s) > 0$ and $\omega_A$,
otherwise the system admits modes that grow like $e^{st}$ where
$\re(s)$ can be arbitrarily large, and the system is ill posed
\cite{KL-Book, GKO-Book}.

Performing the Laplace-Fourier transformation, and assuming trivial
initial data and $\phi=0$, we obtain the following system of ordinary
differential equations
\begin{eqnarray}
s E_x &=& \alpha\partial_x W^A_{\;\; A} 
 + i\omega^A\left[ W_{Ax} - (1+\alpha)W_{xA} \right],\\
s E_B &=& \partial_x Q_B + i\omega^A W_{AB} - i(1+\alpha)\omega^A W_{BA} 
 + i\alpha\omega_B(W_{xx} + W^A_{\;\; A}),\\
s W_{xx} &=& \left(1 + \beta/2 \right)\partial_x E_x + \frac{i\beta}{2}\;\omega^A E_A\; ,\\
s W_{xA} &=& \partial_x E_A\; ,\\
s W_{Ax} &=& i\omega_A E_x\; ,\\
s W_{AB} &=& i\omega_A E_B + \frac{\beta}{2}\; \delta_{AB}\left( \partial_x E_x + i\omega^C E_C \right),
\end{eqnarray}
where we have introduced $Q_A = W_{xA} - (1+\alpha)W_{Ax}$. From this
we can eliminate the variables with zero speed since their evolution
equation becomes algebraic:
\begin{eqnarray}
W_{Ax} &=& \frac{i\omega_A}{s} E_x\; , 
\label{Eq:Zero1}\\
\chi_{AB} &=& \frac{i}{s}\left[ \omega_A E_B - \frac{1}{2}\; \delta_{AB}\omega^C E_C \right],
\label{Eq:Zero2}\\
\kappa &=& -\frac{i}{2s}( 2 + 3\beta) \omega^A E_A\; .
\label{Eq:Zero3}
\end{eqnarray}
Suppose $|\omega| = \sqrt{\delta^{AB}\omega_A\omega_B} \neq 0$. Let
$\hat{\omega}_A = \omega_A/|\omega|$, and let $\hat{\eta}_A$ be a unit
two-vector which is orthogonal to $\omega_A$. Introduce the rescaled
variables $\zeta = s/|\omega|$ and $\xi = |\omega| x$, and define
$E_\omega = E_A\hat{\omega}^A$, $E_\eta = E_A\hat{\eta}^A$, $Q_\omega
= Q_A\hat{\omega}^A$, $Q_\eta = Q_A\hat{\eta}^A$.  In terms of these
variables, the evolution system decouples into the following two
blocks
\begin{equation}
\partial_\xi\left( \begin{array}{c} E_\eta \\ Q_\eta \end{array} \right)
 = \left( \begin{array}{cc} 0 & \zeta \\ \zeta + \frac{1}{\zeta} & 0 \end{array} \right)
\left( \begin{array}{c} E_\eta \\ Q_\eta \end{array} \right),
\end{equation}
\begin{equation}
\partial_\xi\left( \begin{array}{c} E_x \\ Q \\ E_\omega \\ Q_\omega \end{array} \right)
 = \left( \begin{array}{cccc} 0 & \frac{\zeta}{\alpha\beta} & -i\frac{1+\beta}{\beta} & 0 \\
 \zeta + \frac{1-(1+\alpha)^2}{\zeta} & 0 & 0 & i(1+\alpha) \\
 i(1+\alpha) & 0 & 0 & \zeta \\
 0 & -i\frac{1+\beta}{\beta} & \zeta - \frac{\alpha}{\beta}\frac{2\beta+1}{\zeta} & 0 \end{array} \right)
\left( \begin{array}{c}  E_x \\ Q \\ E_\omega \\ Q_\omega \end{array} \right),
\end{equation}
where $Q = \alpha W^A_{\;\; A}$. The first matrix has the eigenvalues
$\pm\lambda$, where $\lambda = \sqrt{\zeta^2 + 1}$. The second matrix
has the eigenvalues $\pm\lambda$ and $\pm\mu$, where $\mu =
\sqrt{\zeta^2/(\alpha\beta) + 1}$ (the sign of the square root is
chosen such that for $\re(\zeta) > 0$, $\re(\lambda) > 0$ and
$\re(\mu) > 0$). The solutions which are bounded as $x
\rightarrow\infty$ are given by
\begin{equation}
\left( \begin{array}{c} E_\eta \\ Q_\eta \end{array} \right)
 = \sigma_0\left( \begin{array}{c} -\zeta \\ \lambda  \end{array} \right) e^{-\lambda\xi},
\label{Eq:Sol1}
\end{equation}
\begin{equation}
\left( \begin{array}{c} E_x \\ Q \\ E_\omega \\ Q_\omega \end{array} \right)
 = \sigma_1\left( \begin{array}{c} 1 \\ \frac{\alpha\lambda}{\zeta} \\ -i\lambda \\ i\frac{\zeta^2-\alpha}{\zeta} \end{array} \right)
e^{-\lambda\xi} 
 + \sigma_2\left( \begin{array}{c} 1 \\ -\frac{\zeta^2-\alpha}{\zeta\mu} \\ -\frac{i}{\mu} \\ -i\frac{\alpha}{\zeta} \end{array} \right)
e^{-\mu\xi},
\label{Eq:Sol2}
\end{equation}
where $\sigma_0$, $\sigma_1$, $\sigma_2$ are integration constants. A
necessary condition for the system to be well posed is that the
boundary conditions must determine these constants uniquely. Using the
relations (\ref{Eq:Zero1},\ref{Eq:Zero2},\ref{Eq:Zero3}) and the
previous notation the boundary conditions
(\ref{Eq:CPBCflat},\ref{Eq:Phys}) yield
\begin{eqnarray}
&& (1-c)[ i\alpha E_\omega - \zeta Q ] 
 + (1+c)\sqrt{\alpha\beta}\left[ \left(\zeta - \frac{\alpha}{\zeta} \right) E_x + i Q_\omega \right] = 0,\\
&& (1-d)E_\omega - (1+d)\left( Q_\omega + i\frac{\alpha-a}{\zeta}\, E_x \right) = g_\omega\; ,\\
&& (1-d)E_\eta - (1+d) Q_\eta = g_\eta\; ,
\end{eqnarray}
where for simplicity we have assumed that $S_A^B = d\,\delta_A^B$ is
diagonal. Plugging into this the general decaying solution
(\ref{Eq:Sol1},\ref{Eq:Sol2}) we obtain the following equations
\begin{equation}
\left[ d_-\zeta - d_+\lambda \right]\sigma_0 = g_\eta\; ,
\end{equation}
\begin{equation}
\left( \begin{array}{cc} d_-\zeta\lambda - d_+(\zeta^2-a)  &  d_-\zeta + a d_+\mu \\
  0 & c_-\zeta - \sqrt{\alpha\beta}c_+\, \mu \end{array} \right) 
\left( \begin{array}{c} \sigma_1 \\ \sigma_2/\mu \end{array} \right)
 = \left( \begin{array}{c} -i\zeta g_\omega \\ 0 \end{array} \right) ,
\end{equation}
where $c_\pm = c \pm 1$, $d_\pm = d \pm 1$. In order to analyze in
what cases these equations uniquely determine the constants
$\sigma_0$, $\sigma_1$ and $\sigma_2$ we use the following

\begin{lemma}
\label{Lem:Func}

Let $P > 0$, $A,B \in \Real$, $(A,B) \neq (0,0)$, and consider the
function
\begin{displaymath}
\psi: \{ \re(\zeta) > 0 \} \rightarrow \Complex,
\psi(\zeta) = A\,\zeta - B\sqrt{\zeta^2 + P^2}, 
\end{displaymath}
where we choose the branch such that $\re(\sqrt{\zeta^2 + P^2}) > 0$
for $\re(\zeta) > 0$.

Then, $\psi$ has zeroes if and only if $A > B > 0$ or $A < B < 0$.
Furthermore, $|\psi|$ is uniformly bounded away from zero if and only if $A\cdot B < 0$.
\end{lemma}

{\bf Proof}:
If $B=0$ the Lemma is trivial. So let $B\neq 0$ and rescale $\zeta$,
$\psi$ and $A$ such that $\psi(\zeta) = A\,\zeta - \sqrt{\zeta^2 + 1}$.
Suppose $\psi(\zeta) = 0$ for some $\re(\zeta) > 0$. Then it follows
that $(A^2-1)\zeta^2 = 1$ and thus $A^2 > 1$.  Therefore, $\zeta =
1/\sqrt{A^2-1}$ and $\sqrt{\zeta^2 + 1} = |A|/\sqrt{A^2-1}$. This
satisfies $\psi(\zeta) = 0$ if and only if $A > 1$ is positive. This
proves the first statement of the Lemma.

In order to prove the second assertion of the Lemma we restrict
ourselves to the case $A \leq 1$ since otherwise $\psi$ has zeroes.
Notice first that for large $|\zeta|$, $\psi(\zeta) = (A-1)\zeta +
O(\zeta^{-1})$, hence $|\psi|$ is not bounded away from zero if $A=1$.
So let $A < 1$, and let $\zeta_j$ be a sequence with $\re(\zeta_j) >
0$ such that $\psi(\zeta_j) \rightarrow 0$. This sequence is bounded
for $j$ large enough since otherwise $A=1$. Therefore, there is a
subsequence that converges to some $\zeta^*\in\Complex$.  We must have
$\re(\zeta^*) = 0$ since otherwise $\zeta^*$ is a zero of $\psi$. So
we must determine in what cases zero lies on the boundary of the image
of $\psi$ in $\Complex$. For $\zeta = i\cos\alpha$, $0 < \alpha <
\pi$, we have $\sqrt{\zeta^2+1} = \sin\alpha$. Next, let $\gamma > 0$,
$\varepsilon > 0$ and $\zeta = i\cosh\gamma + \varepsilon$. Then
\begin{displaymath}
\sqrt{\zeta^2+1} = \sqrt{-\sinh^2\gamma + 2i\cosh(\gamma)\varepsilon + \varepsilon^2}
 = +i\sinh\gamma + \coth(\gamma)\varepsilon + O(\varepsilon^2),
\end{displaymath}
where we have chosen the sign such that $\re(\sqrt{\zeta^2 + 1}) > 0$
for $\varepsilon$ small enough. Therefore, the boundary of the image
of $\psi$ can be parametrized by
\begin{eqnarray}
\psi(i\cos\alpha) &=& -\sin\alpha + iA\cos\alpha,
\qquad 0 < \alpha < \pi,\\
\psi(\pm i\cosh\gamma) &=& \pm i(A\cosh\gamma - \sinh\gamma),
\qquad \gamma \geq 0.
\end{eqnarray}
We see that the boundary contains zero only if $0 \leq A \leq 1$.
Therefore, $|\psi|$ is bounded away from zero if and only if $A < 0$.
\qed

We first apply the Lemma to the case $a=0$ and $\alpha\beta > 0$. It
follows that the $\sigma_0$, $\sigma_1$ and $\sigma_2$ are uniquely
determined if and only if $-1 \leq c$ and $-1 \leq d \leq 1$.
Therefore, the conditions $-1 \leq c$, $|d| \leq 1$ are {\em
necessary} for the well posedness of the initial-boundary value
problem. This justifies the restriction on the matrix $S = (S_A^{\;\;
B})$ to satisfy $S^\dagger S \leq 1$. Next, we impose the Kreiss
determinant condition \cite{Kreiss, KL-Book, GKO-Book} which is
stronger and requires that the constants $\sigma_0$ and $\sigma_1$ can
be bounded by the boundary data with a bound that is independent on
$\omega$ and $\zeta$. It follows from Lemma \ref{Lem:Func} that this
condition is satisfied if and only if $|d| < 1$. This immediately
implies that the problem is well posed in $L^2$ if the initial data is
trivial, since one can estimate the integration constants $\sigma_0$,
$\sigma_1$ by the boundary data (see Lemma 8.4.3. in \cite{KL-Book})
while $\sigma_2 = 0$ if $-1 \leq c$.

On the other hand, suppose that $d=0$ and $a = \alpha$ which
corresponds in setting to zero the transversal ingoing characteristic
fields. The resulting initial-boundary value problem has been
integrated in Ref. \cite{LS-FatMax} by numerical means for different
values of the parameters $\alpha$ and $\beta$, subject to the
restriction $\alpha\beta > 0$ (in the notation of \cite{LS-FatMax},
$\alpha = -\gamma_1/2$, $\beta = -2\gamma_2$). In this case,
$d_-\zeta\lambda - d_+(\zeta^2-a) = -[ (\lambda+\zeta)^2 - 2\alpha -
1]/2$ and since the map $\zeta \mapsto (\lambda + \zeta)^2$ maps
$\re(\zeta) > 0$ onto the outside of the unit disk minus the negative
real axis, the system is ill posed if $2\alpha + 1 > 1$, that is, if
$\alpha > 0$. Indeed, the numerical results of Ref. \cite{LS-FatMax}
exhibit instabilities in those cases.

\section{Energy estimates}
\label{sect:EE}

After the preliminary investigations in the previous section, we
return to the case where $\Omega\in\Real^3$ is a bounded domain with
$C^\infty$ boundary $\partial\Omega$, and derive some a priori
estimates in this section. For simplicity, we only consider the
source-free case here, where $\rho=0$, $J_i=0$; the generalization to
the inhomogeneous case is discussed at the end of next section.  We
assume that we are given a smooth function $u = (A_i, E_i, W_{ij})$
which satisfies the evolution equations
(\ref{Eq:Af},\ref{Eq:Ef},\ref{Eq:Wf}), and the boundary conditions
(\ref{Eq:CPBC},\ref{Eq:PoyntingBC}). The goal of this section is to
show that this solution satisfies the estimates of Theorem
\ref{Thm:Main}. These estimates are then used in the next section in
order to prove existence with semigroup methods.

We start with estimates for the gauge invariant quantities $E_i$ and
$F_{ij} = W_{ij} - W_{ji}$ (i.e. the electric and magnetic fields),
and the constraint variables $C$, $C_k$, $C^{(W)}_{ij}$, and then
estimate the gauge dependent quantities $A_i$ and the symmetric part
of $W_{ij}$. It is the estimate for the latter quantities that
requires a specific gauge condition, since the counterexample
presented in section \ref{sect:MP} shows that the simple condition
$\phi = 0$ does not allow to estimate the symmetric part of $W_{ij}$
in $L^2$. Assume first that $\alpha=\beta=0$. In this case, the
evolution system (\ref{Eq:Af},\ref{Eq:Ef},\ref{Eq:Wf}) is only weakly
hyperbolic. However, in this case, one can replace $W_{ij}$ by
$F_{ij}$ by taking the antisymmetric part of Eq. (\ref{Eq:Wf}) and
discarding its symmetric part. Consider the boundary conditions
(\ref{Eq:PoyntingBC}):
\begin{equation}
w_i^{(+)} = S_i^{\;\; j} w_j^{(-)} + g_i\; , \qquad
w_i^{(\pm)} \equiv H_i^{\;\; j} E_j \pm n^k F_{ki}\; ,
\label{Eq:PoyntingBCbis}
\end{equation}
where we recall that $S_i^{\;\; j}$ is a smooth matrix-valued function
on $\partial\Omega$ with the properties that $S_i^{\;\; j} n^i = 0$,
$S_i^{\;\; j} n_j = 0$ and $H^{ij} S_i^{\;\; r} S_j^{\;\; s} \leq
H^{rs}$ if $g_i = 0$ and $H^{ij} S_i^{\;\; r} S_j^{\;\; s} \leq \delta
H^{rs}$ for a $\delta < 1$ otherwise. The physical energy is defined
by
\begin{equation}
{\cal E}_{phys} = \frac{1}{2}\int_{\Omega} \left( E^j E_j + \frac{1}{2} F^{ij} F_{ij} \right) d^3 x,
\end{equation}
Taking a time derivative, using Eqs. (\ref{Eq:Ef},\ref{Eq:Wf})
and Gauss' theorem, we obtain
\begin{eqnarray}
\frac{d}{dt} {\cal E}_{phys} &=& \int_{\Omega} \left( E^j\nabla^i F_{ij} + F^{ij} \nabla_i E_j \right) d^3 x
 = \int_{\partial\Omega} n^i E^j F_{ij} dS
\nonumber\\ 
 &=& \frac{1}{4} \int_{\partial\Omega} H^{ij} \left( w_i^{(+)} w_j^{(+)} -  w_i^{(-)} w_j^{(-)} \right) dS,
\nonumber
\end{eqnarray}
where $dS$ denotes the surface element on $\partial\Omega$.
If the boundary conditions are homogeneous ($g_i = 0$) it follows
immediately that the boundary integral is negative or zero, and
we have an estimate for ${\cal E}_{phys}$. If the boundary data
is nonzero, we have the estimate \cite{KL-Book}
\begin{equation}
H^{ij} \left( w_i^{(+)} w_j^{(+)} -  w_i^{(-)} w_j^{(-)} \right)
 \leq -\varepsilon H^{ij} \left( w_i^{(+)} w_j^{(+)} +  w_i^{(-)} w_j^{(-)} \right) 
 + K H^{ij} g_i g_j\; ,
\end{equation}
for some constants $\varepsilon = \varepsilon(\delta) > 0$, $K =
K(\delta) > 0$ from which we conclude
\begin{equation}
{\cal E}_{phys}(t) \leq {\cal E}_{phys}(0) + a\int_0^t \| \bbg(s) \|_{L^2(\partial\Omega)}^2 ds
\end{equation}
for some constant $a = a(\delta) > 0$.

When $\alpha\beta > 0$, we obtain
\begin{eqnarray}
\frac{d}{dt} {\cal E}_{phys} &=& \int_{\Omega} \left( E^j(\nabla^i F_{ij} + \alpha C_j) + F^{ij} \nabla_i E_j \right) d^3 x
\nonumber\\
 &\leq& \alpha \int_{\Omega} E^j C_j\, d^3 x + a\int_{\partial\Omega} H^{ij} g_i g_j\, dS
\nonumber
\end{eqnarray}
which contains the additional term $\int_{\Omega} E^j C_j\, d^3 x$
which cannot be estimated by ${\cal E}_{phys}$ alone. However, from
the evolution system for the constraints,
Eqs. (\ref{Eq:Cijf},\ref{Eq:Cf},\ref{Eq:Ckf}), and the
constraint-preserving boundary condition (\ref{Eq:CPBC}),
\begin{equation}
V^{(+)} = c\; V^{(-)}, \qquad
V^{(\pm)} \equiv C \mp \frac{\sqrt{\alpha\beta}}{\beta}\, n^k C_k\; ,
\label{Eq:CPBCbis}
\end{equation}
where $c$ is a smooth function on $\partial\Omega$ with $|c|\leq 1$,
we can estimate the norm
\begin{equation}
{\cal E}_{cons} = \frac{1}{2}\int_{\Omega} \left( n_1 C^{(W)ij} C^{(W)}_{ij} 
 + \tau^2 C^2 + \tau^2\frac{\alpha}{\beta}\, C^k C_k \right) d^3 x.
\end{equation}
Here, $\tau > 0$ is a fixed constant with dimension of length and
$n_1$ is a positive constant. We obtain
\begin{eqnarray}
\frac{d}{dt} {\cal E}_{cons} &=& -\int_{\Omega} \left( n_1\frac{\beta}{2}\, h^{ij} C^{(W)}_{ij} C 
 + \alpha\tau^2\left( C\nabla^k C_k + C^k\nabla_k C \right) \right) d^3 x
\nonumber\\
 &=& -n_1\frac{\beta}{2} \int_{\Omega} h^{ij} C^{(W)}_{ij} C\, d^3 x - \alpha\tau^2\int_{\partial\Omega} n^k C C_k dS
\nonumber\\
 &\leq& \frac{r}{2\tau} \int_{\Omega} \left( \frac{4\alpha}{3\beta} C^{(W)ij} C^{(W)}_{ij}  + \tau^2 C^2 \right) d^3 x,
\nonumber
\end{eqnarray}
where we have used Schwarz' inequality and the inequality $(h^{ij}
C^{(W)}_{ij})^2 \leq 3 C^{(W)ij} C^{(W)}_{ij}$ in the last step, and
where we have set $n_1 = 4\alpha/(3\beta)$ and $r =
\sqrt{\alpha\beta}$. On the other hand, using Schwarz' inequality
again, we can estimate
\begin{displaymath}
\alpha\int_{\Omega} E^j C_j\, d^3 x \leq \frac{r}{2\tau} \int_{\Omega} 
\left\{ E^j E_j + \tau^2\frac{\alpha}{\beta} C^j C_j \right) d^3 x.
\end{displaymath}
Therefore, we obtain an estimate for the energy norm
\begin{equation}
{\cal E} = {\cal E}_{phys} + {\cal E}_{cons}\; ,
\end{equation}
namely,
\begin{equation}
\frac{d}{dt} {\cal E} \leq \frac{r}{\tau}\, {\cal E} + a\int_{\partial\Omega} H^{ij} g_i g_j\, dS ,
\end{equation}
which shows that we can estimate the gauge-invariant quantities $E_j$,
$F_{ij}$, $C$, $C_k$, $C^{(W)}_{ij}$. Notice that we can choose $\tau$
arbitrary large, and thus we can make the exponential growth rate in
the bound as small as we like. What is missing are estimates for the
magnetic potential $A_j$ and the symmetric part of $W_{ij}$. Such
estimates depend on the gauge choice for $\phi$. We have seen that the
problem is not well posed in $L^2$ if we choose the temporal gauge
$\phi = 0$. An $L^2$ estimate can be obtained if we impose the
following gauge choice instead:
\begin{equation}
\Delta\phi = -\nabla^k E_k, \qquad
n^k\nabla_k\phi = -n^k E_k\quad \hbox{on $\partial\Omega$},
\label{Eq:NewGauge}
\end{equation}
where $\Delta = \nabla^k\nabla_k$. The boundary condition on $\phi$
implies that on $\partial\Omega$, $\partial_t (n^k A_k) = n^k E_k +
n^k \nabla_k\phi = 0$. This implies that $\left. n^k A_k
\right|_{\partial\Omega} = 0$ provided the initial data satisfies this
condition. Next, we notice that we can estimate $W$ since $\partial_t
W = -(1+ 3\beta/2)C + \Delta\phi = -3\beta C$. Since we control
$C^{(W)}_{ij} \equiv W_{ij} - \nabla_i A_j$ we can also estimate
$\nabla^i A_i$. On the other hand, using Gauss' theorem twice,
\begin{eqnarray}
\int_{\Omega} (\nabla_i A^i)(\nabla_j A^j) d^3 x 
 &=& \int_{\Omega} (\nabla_j A^i)(\nabla_i A^j) d^3 x
 + \int_{\partial\Omega} n^i\left( A_i\nabla_j A^j - A^j\nabla_j A_i \right) dS
\nonumber\\
&=& \int_{\Omega} (\nabla_j A^i)(\nabla_i A^j) d^3 x
 + \int_{\partial\Omega} \kappa_{ij} A^i A^j dS,
\label{Eq:Asym}
\end{eqnarray}
where in the last term, $\kappa_{ij} = (h_{ik} - n_i n_k)\nabla^k n_j$
denotes the extrinsic curvature of the boundary $\partial\Omega$, and
where we have used the fact that $A^i n_i = 0$ on
$\partial\Omega$. Assuming that $\kappa_{ij}$ is positive
semi-definite at each point of $\partial\Omega$\footnote{This
condition can probably be weakened by using the trace theorems.}, we
obtain the inequality
\begin{equation}
\int_{\Omega} \left( 2\nabla^{[i} A^{j]}\cdot\nabla_{[i} A_{j]} + \nabla_i A^i\cdot\nabla_j A^j \right) d^3 x 
 \geq \int_{\Omega} \nabla^i A^j\cdot \nabla_i A_j\, d^3 x
\label{Eq:AsymEst}
\end{equation}
which allows us to estimate all spatial derivatives of $A_j$; and in particular
the symmetric part of $W_{ij}$ since we control $C^{(W)}_{ij}$.

Finally, we show how to estimate $A_i$: Because of
Eq. (\ref{Eq:NewGauge}) we have
\begin{eqnarray}
&& \int_{\Omega} (E^i + 2\nabla^i\phi)(E_i + 2\nabla_i\phi) d^3 x 
 = \int_{\Omega} \left\{ E^i E_i + 4E^i\nabla_i\phi + 4\nabla^i\phi \nabla_i\phi \right\} d^3 x
\nonumber\\
&& = 4\int_{\partial\Omega} \phi\, n^i\left( E_i + \nabla_i\phi \right) dS 
   + \int_{\Omega} \left\{ E^i E_i - 4\phi\left( \nabla^i E_i + \Delta\phi \right) \right\} d^3 x
\nonumber\\
&& = \int_{\Omega} E^i E_i\; d^3 x,
\label{Eq:nablaphiEst}
\end{eqnarray}
so we can estimate $E_i + 2\nabla_i\phi$ and thus also $\partial_t A_i
= E_i + \nabla_i\phi$ since we have an estimate for $E_i$. The net
result is the a priori estimate
\begin{equation}
\| u(t) \|_{L^2(\Omega)}^2 + \tau^2 \| v(t) \|_{L^2(\Omega)}^2 
 \leq a\, e^{b\,t/\tau} \left[ \| u(0) \|_{L^2(\Omega)}^2 + \tau^2 \| v(0) \|_{L^2(\Omega)}^2 
 + \int_0^t \| \bbg(s) \|_{L^2(\partial\Omega)}^2 ds \right],
\end{equation}
where $a > 0$, $b > 0$ are two constants and, $u = (\tau^{-1} A_i,
E_j, W_{ij})$, $v = (\tau^{-1} C^{(W)}_{ij}, C, C_k)$. This implies the
estimate in Theorem \ref{Thm:Main} in the absence of source functions.

\section{Existence}
\label{sect:existence}

In this section, we prove well posedness for the initial-boundary
value problem defined by the evolution equations
(\ref{Eq:rhof},\ref{Eq:Af},\ref{Eq:Ef},\ref{Eq:Wf}), the gauge
condition (\ref{Eq:EllipticGauge}) and the constraint-preserving
boundary conditions (\ref{Eq:CPBC},\ref{Eq:PoyntingBC}). The idea is
to represent the problem as an abstract Cauchy problem
\begin{displaymath}
\frac{d}{dt} u(t) = {\cal A} u(t),\qquad
u(0) = u_0\in H,
\end{displaymath}
where ${\cal A}: D({\cal A}) \subset H \rightarrow H$ is a linear
operator on an appropriate Hilbert space $H$. This operator is
basically given by the right-hand side of the evolution equations, and
we will define its domain $D({\cal A})$ to be the space of smooth
functions satisfying the boundary conditions with homogeneous boundary
data.  We then show that ${\cal A}$ has a unique extension $\bar{\cal
A}$ and that this extension generates a strongly continuous semigroup,
which we write formally as $P(t) = \exp(t{\cal A})$. Given initial
data $u_0$ in the domain $D(\bar{\cal A})$ of the extension, the
solution to the abstract Cauchy problem is given by
$u(t)=P(t)u_0$. The semigroup properties imply that $\| u(t) \| \leq
a\exp(bt) \| u_0 \|$ for some constants $a > 0$, $b > 0$ which are
independent of $u_0$; thus the problem is well posed. In the
following, we assume that the coupling functions $c$ and $S_i^{\;\;
j}$ that appear in the boundary conditions are time-independent, lie
in $C^\infty(\partial\Omega)$ and satisfy $|c| < 1$, $H^{ij} S_i^{\;\;
r} S_j^{\;\; s} < H^{rs}$ at each point of $\partial\Omega$.

The Hilbert space is motivated by the energy estimate in the previous
section. Let
\begin{eqnarray}
H_A = \{ \bbA\in L^2(\Omega,\Real^3) :
\hbox{The derivatives $\nabla_j A_i$ exist in the weak}&&
\nonumber\\
\hbox{sense and belong to $L^2(\Omega)$, and on $\partial\Omega$, $n^i A_i = 0$.} \},&&
\nonumber\\
H_E = \{ \bbE\in L^2(\Omega,\Real^3) :
\hbox{The derivative $\nabla^i E_i$ exists in the weak}&&
\nonumber\\
\hbox{sense and belongs to $L^2(\Omega)$.} \},&&
\nonumber\\
H_W = \{ \bbW\in L^2(\Omega,\Real^9) :
\hbox{The derivatives $\nabla_j W - \nabla^i W_{ji}$ exist in the weak}&&
\nonumber\\
\hbox{sense and belong to $L^2(\Omega)$.} \}.&&
\nonumber
\end{eqnarray}
Here, $C = -\nabla^i E_i$ exists in the weak sense and belongs to
$L^2(\Omega)$ means that $C\in L^2(\Omega,\Real)$ has the property
that
\begin{displaymath}
\int_\Omega C\,\phi\, dx = \int_\Omega E^j \nabla_j\phi\, dx 
\end{displaymath}
for all test functions $\phi\in C_0^\infty(\Omega,\Real)$. Similar
definitions apply for the weak derivatives $\nabla_j A_i$ and
$\nabla_j W - \nabla^i W_{ji}$. We introduce the following scalar
products on $H_A$, $H_E$ and $H_W$:
\begin{eqnarray}
\sprod{\bbA}{\bbB}_A &=& \int_\Omega \left( \tau^{-2} A^j B_j + \nabla^i A^j\cdot\nabla_i B_j \right) d^3 x,
\nonumber\\
\sprod{\bbE}{\bbF}_E &=& \int_\Omega \left( E^j F_j + \tau^2 \nabla^i E_i\cdot\nabla^j F_j \right) d^3 x,
\nonumber\\
\sprod{\bbW}{\bbV}_W &=& \int_\Omega \left( W^{ij} V_{ij} 
 + \tau^2 (\nabla^i W - \nabla_j W^{ij})(\nabla_i V - \nabla^k V_{ik}) \right) d^3 x,
\nonumber
\end{eqnarray}
where $\bbA,\bbB\in H_A$, $\bbE,\bbF\in H_E$, and $\bbW,\bbV\in H_W$.
Notice that the requirement $n^i A_i = 0$ on $\partial\Omega$ makes
sense because of the trace theorems. The following results are
standard.

\begin{lemma}
\label{Lem:Hilbert}
The spaces $(H_A,\sprod{.}{.}_A)$, $(H_E,\sprod{.}{.}_E)$ and
$(H_W,\sprod{.}{.}_W)$ are Hilbert spaces.
\end{lemma}

\begin{lemma}
\label{Lem:Dense}
Denote by $C^\infty(\bar{\Omega},\Real^m)$ the class of functions
$\bar{\Omega} \rightarrow \Real^m$ which are the restriction of a
smooth function $C^\infty(\Real^3,\Real^m)$ on $\bar{\Omega}$. Then,\\
$C^\infty(\bar{\Omega},\Real)$ is dense in $L^2(\Omega,\Real)$.\\
$\{ \bbA\in C^\infty(\bar{\Omega},\Real^3): \hbox{$n^i A_i=0$ on
$\partial\Omega$} \}$ is dense in $H_A$.\\
$C^\infty(\bar{\Omega},\Real^3)$ is dense in $H_E$.\\
$C^\infty(\bar{\Omega},\Real^9)$ is dense in $H_W$.
\end{lemma}

We define the total Hilbert space $H = L^2(\Omega,\Real) \times H_A
\times H_E \times H_W$ with scalar product
\begin{eqnarray}
\sprod{u}{\bar{u}}_H &=& \int_{\Omega} \left( \tau^2\rho^2 + \tau^{-2} A^i\bar{A}_i + E^i\bar{E}_i + 2W^{[ij]}\bar{W}_{[ij]} 
  + \frac{\alpha}{3\beta} W\bar{W} \right.
\nonumber\\
 && \left.\qquad +\, \frac{\alpha}{3\beta} C^{(W)ij}\bar{C}^{(W)}_{ij} 
 + \tau^2 C\bar{C} + \tau^2\frac{\alpha}{\beta}\, C^k\bar{C}_k \right) d^3 x,
\end{eqnarray}
where $u = (\rho,A_i,E_i,W_{ij})$, $C^{(W)}_{ij} = W_{ij} - \nabla_i
A_j$, $C = \rho - \nabla^k E_k$, $C_k = \nabla_k W - \nabla^j
W_{kj}$. The estimate (\ref{Eq:AsymEst}) implies that the norm induced
by $\sprod{.}{.}_H$ is equivalent to the one induced by the scalar
product constructed from $\sprod{.}{.}_{L^2}$, $\sprod{.}{.}_A$,
$\sprod{.}{.}_E$ and $\sprod{.}{.}_W$. Also, the Hilbert space $H$ is
topologically equivalent to the space ${\cal H}'$ defined in section
\ref{sect:MP}. Before we define the operator ${\cal A}$ on the Hilbert
space $H$, we need the following result from elliptic theory (see, for
example \cite{Taylor}):

\begin{lemma}
Let $F\in C^\infty(\bar{\Omega})$, and $g\in
C^\infty(\partial\Omega)$. Then, the Neumann problem
\begin{eqnarray}
&& \Delta u = F \qquad \hbox{on $\Omega$},\\
&& n^k\nabla_k u = g \qquad \hbox{on $\partial\Omega$},
\end{eqnarray}
has a solution $u\in C^\infty(\bar{\Omega})$ if and only if
\begin{equation}
\int_{\Omega} F\, d^3 x = \int_{\partial\Omega} g\, dS.
\end{equation}
The solution is unique up to an additive constant.
\end{lemma}

This lemma allows to solve the boundary value problem
(\ref{Eq:NewGauge}) since according to Gauss' theorem $\int_{\Omega}
\nabla^k E_k\, d^3 x = \int_{\partial\Omega} n^k E_k dS$. We write
$\phi = -\Delta^{-1}\nabla^k E_k + const$ for its solution. We define
the linear operator ${\cal A}: D({\cal A})\subset H \rightarrow H$ on
the Hilbert space $H$ by
\begin{eqnarray}
D({\cal A}) = \{ (\rho,\bbA,\bbE,\bbW)\in C^\infty(\bar{\Omega},\Real^{16})
: \hbox{On $\partial\Omega$ we have}\, n^i A_i = 0, \; &&
\nonumber\\
  H_i^{\;\; j} E_j + a_i^{\;\; j} n^k(W_{kj} - W_{jk}) = 0, 
 \nabla^k E_k - \rho + b_0 \frac{r}{\beta}\, n^i(\nabla_i W - \nabla^j W_{ij}) = 0\}, &&
\label{Eq:DefDA}
\end{eqnarray}
where ${\bf a} = (a_i^{\;\; j}) = {\bf H}({\bf H} - {\bf S})^{-1}
({\bf H} + {\bf S})$ and $b_0 = (1+c)/(1-c)$, and
\begin{equation}
{\cal A}\left( \begin{array}{l} \rho \\ A_j \\ E_j \\ W_{ij} \end{array} \right) = 
\left( \begin{array}{l} 0 \\ E_j - \nabla_j\Delta^{-1}\nabla^i E_i \\ 
\nabla^i W_{ij} - (1+\alpha) \nabla^i W_{ji} + \alpha \nabla_j W \\ 
\nabla_i E_j + \frac{\beta}{2}\, h_{ij} \nabla^k E_k 
 - \nabla_i\nabla_j\Delta^{-1}\nabla^k E_k -  \frac{\beta}{2}\, h_{ij} \rho
 \end{array} \right).
\label{Eq:DefA}
\end{equation}

The existence proof relies on the following three propositions, which
imply by the Lumer-Phillips theorem \cite{Pazy} that ${\cal A}$ is
closable and that its closure is the generator of a strongly
continuous semigroup.

\begin{proposition}
\label{Prop:Dense}
$D({\cal A})$ is dense in $H$.
\end{proposition}

\begin{proposition}
\label{Prop:QuasiDiss}
The operator ${\cal A}: D({\cal A}) \rightarrow H$ defined in
(\ref{Eq:DefDA},\ref{Eq:DefA}) is quasi-dissipative. That is, there is
a constant $b$ such that
\begin{equation}
\re\sprod{u}{{\cal A}u}_H \leq \frac{b}{\tau}\, \sprod{u}{u}_H
\end{equation}
for all $u\in D({\cal A})$.
\end{proposition}

\begin{proposition}
\label{Prop:DenseCond}
$(\lambda-{\cal A})(D({\cal A}))$ is dense in $H$ for $\lambda > 0$
sufficiently large.
\end{proposition}

\begin{theorem}[Well posedness in the homogeneous case]
\label{Thm:Homo}
The linear operator ${\cal A}: D({\cal A}) \rightarrow H$ is closable
and its closure $\bar{\cal A}$ is the generator of a strongly
continuous semigroup $P(t)$ in $H$. Given initial data $u_0\in
D(\bar{\cal A})$, the map $\Real_+ \rightarrow D(\bar{\cal A})$, $t
\mapsto P(t)u_0$ is strongly differentiable and satisfies
\begin{equation}
\frac{d}{dt} u(t) = \bar{\cal A} u(t), \qquad t > 0,
\end{equation}
and so gives a solution to the constrained evolution system
(\ref{Eq:rhof},\ref{Eq:Af},\ref{Eq:Ef},\ref{Eq:Wf}), (\ref{Eq:Gauss},\ref{Eq:Cij})
with $J_j=0$ and homogeneous boundary data, $g_i=0$. This
solution obeys the estimate (\ref{Eq:energy_estimate1}).
\end{theorem}

We will prove later that this solution also obeys the estimate (\ref{Eq:energy_estimate2}).

\subsection{Proof of Proposition \ref{Prop:Dense}}

The proof of Proposition \ref{Prop:Dense} is based on Lemma
\ref{Lem:Dense} and

\begin{lemma}
\label{Lem:SmallEBC}
Let $G\in C^\infty(\partial\Omega,\Real)$, $\bbF\in
C^\infty(\partial\Omega,\Real^3)$ with $F_i n^i = 0$, and let
$\varepsilon > 0$. There exists $\bbE\in
C^\infty(\bar{\Omega},\Real^3)$ such that $\| \bbE \|_E < \varepsilon$
and such that on $\partial\Omega$,
\begin{displaymath}
H_i^{\; j} E_j = F_i\, , \qquad
\nabla^k E_k = G,
\end{displaymath}
where $n_i$ denotes the unit outward normal to $\Omega$, and $H_i^{\;
j} = \delta_i^{\; j} - n_i n^j$ is the projection operator on the
tangent space of $\partial\Omega$.
\end{lemma}

{\bf Proof}:
We construct $\bbE$ in the following way: First, extend $n_i$ to a
neighborhood of $\partial\Omega$ in $\Omega$ by shooting geodesics
through $n_i$ at each point of $\partial\Omega$ (such that
$n^i\nabla_i n_k = 0$). Denote by $s$ the affine parameter which is
such that $\nabla_i s = n_i$ and $s=0$ on $\partial\Omega$. We
consider a neighborhood $U_\delta$ of $\Omega$ which is spanned by
$s\in(-\delta,0)$ for some $\delta > 0$.  Next, extend $G$ and $F_i$
inside this neighborhood by solving the ordinary differential
equations
\begin{eqnarray}
&& \nabla_n G + \kappa G = 0,
\nonumber\\
&& \nabla_n F_i + \kappa F_i = \kappa_{ji} F^j,
\nonumber
\end{eqnarray}
where $\kappa_{ij} = \nabla_i n_j$ is the extrinsic curvature of the
surfaces $s=const$. Notice that the second equation implies that
$\nabla_n(n^i F_i) = -\kappa(n^i F_i)$, so $n^i F_i = 0$ on
$U_\delta$. Finally, let $m > 2/\delta$ and define the function
$\psi_m\in C^\infty(\bar{\Omega})$ by $\psi_m = m^{-1}\psi(m s)$,
where $\psi\in C^\infty((-\infty,0],\Real)$ has the following
properties:
\begin{enumerate}
\item[(i)] $\psi(0) = 1$, $\psi(s) = 0$ for all $s\leq -2$.
\item[(ii)] $\psi'(0) = 1$
\item[(iii)] $0 \leq \psi'(s) \leq 1$ for all $s \leq 0$.
\end{enumerate}

Then, we define $E_i = U_i + \varepsilon_{irs}\nabla^r V^s$, where
$U_i = \psi_m n_i G$, $V^s = -\psi_m \varepsilon^{skl} n_k F_l$ and
$\varepsilon_{ijk}$ denotes the natural volume element on $\Omega$. By
construction, we have
\begin{eqnarray}
&& \nabla^i E_i = \nabla^i U_i = \nabla_n\psi_m\cdot G,
\nonumber\\
&& E_i = \psi_m n_i (G - \nabla^j F_j) + \nabla_n\psi_m\cdot F_i\; ,
\nonumber
\end{eqnarray}
so $E_i$ satisfies the boundary conditions. Furthermore,
\begin{eqnarray}
\| \bbE \|_E^2 &=& \int_{U_\delta} \left[ \psi_m^2 (G - \nabla^j F_j)^2 + (\nabla_n\psi_m)^2( F^i F_i + \tau^2 G^2) \right] d^3 x
\nonumber\\
 &\leq& | Vol(U_{2/m}) | \left[ \frac{1}{m^2} \int_{U_\delta} (G - \nabla^j F_j)^2 d^3 x
 + \int_{U_\delta} ( F^i F_i + \tau^2 G^2) d^3 x \right].
\nonumber
\end{eqnarray}
The right-hand side converges to zero as $m\rightarrow\infty$, so the
lemma follows.
\qed

Now let $(\rho,\bbA,\bbE,\bbW)\in H$, and let $\varepsilon > 0$ be
arbitrarily small. According to Lemma \ref{Lem:Dense} there exist
$\tilde{\rho}$, $\tilde{\bbA}, \tilde{\bbE}^{(1)},\tilde{\bbW} \in
C^\infty(\bar{\Omega})$ with $\tilde{A}_n = 0$ on $\partial\Omega$
such that
\begin{displaymath}
\| \rho - \tilde{\rho} \|_{L^2} < \frac{\varepsilon}{5}\; ,\qquad
\| \bbA - \tilde{\bbA} \|_A < \frac{\varepsilon}{5}\; ,\qquad
\| \bbE - \tilde{\bbE}^{(1)} \|_E < \frac{\varepsilon}{5}\; ,\qquad
\| \bbW - \tilde{\bbW} \|_W < \frac{\varepsilon}{5}\; .
\end{displaymath}
Using Lemma \ref{Lem:SmallEBC} we can find $\tilde{\bbE}^{(2)}\in
C^\infty(\bar{\Omega})$ such that $\| \tilde{\bbE}^{(2)} \|_E <
\varepsilon/5$ and such that on $\partial\Omega$,
\begin{eqnarray}
H_i^{\;\; j} \tilde{E}_j^{(2)} &=& -H_i^{\;\; j} \tilde{E}_j^{(1)} 
         - a_i^{\;\; j} n^k(\tilde{W}_{kj} - \tilde{W}_{jk}),
\nonumber\\
\nabla^k\tilde{E}_k^{(2)} &=& -\nabla^k\tilde{E}_k^{(1)} 
     - b_0\frac{r}{\beta}\, n^i(\nabla_i\tilde{W} - \nabla^k\tilde{W}_{ik}) + \tilde{\rho}.
\nonumber
\end{eqnarray}
Therefore, $(\tilde{\rho},\tilde{\bbA},\tilde{\bbE}^{(1)} +
\tilde{\bbE}^{(2)},\tilde{\bbW})\in D({\cal A})$ and
\begin{displaymath}
\| \rho - \tilde{\rho} \|_{L^2} +
\| \bbA - \tilde{\bbA} \|_A + \| \bbE - \tilde{\bbE}^{(1)} - \tilde{\bbE}^{(2)} \|_E + \| \bbW - \tilde{\bbW} \|_W 
 < \varepsilon.
\end{displaymath}
This proves Proposition \ref{Prop:Dense}.

\subsection{Proof of Proposition \ref{Prop:QuasiDiss}}

Proposition \ref{Prop:QuasiDiss} follows almost directly from the
estimates in the previous section, so we only give the main steps
here. Let $u\in D({\cal A})$. Using Gauss' theorem and the
estimate (\ref{Eq:nablaphiEst}), we have, setting $\phi = -\Delta^{-1}\nabla^k E_k$,
\begin{eqnarray}
\sprod{u}{{\cal A}u}_H &=& \int_{\Omega} \left\{ \frac{1}{2\tau^2} A^j(E_j + 2\nabla_j\phi) 
 + \frac{1}{2\tau^2} A^j E_j + \alpha E^j C_j - \frac{\alpha}{2}\, W C 
 - \frac{\alpha}{6} h^{ij} C_{ij}^{(W)} C \right\} d^3 x
\nonumber\\
 &+& \int_{\partial\Omega} \left( 2 n^i E^j W_{[ij]} - \alpha\tau^2 n^i C C_i \right) dS
\nonumber\\
 &\leq& \frac{1}{2\tau}\int_{\Omega} \left\{ \frac{1}{\tau^2} A^j A_j + E^j E_j 
 + r\left[ E^j E_j + \tau^2\frac{\alpha}{\beta}\, C^j C_j 
 + \frac{\alpha}{3\beta}\, W^2 + \frac{3\tau^2}{4}\, C^2
 + \frac{\alpha}{3\beta}\, C^{(W)ij} C^{(W)}_{ij} + \frac{\tau^2}{4}\, C^2 \right\} \right] d^3 x
\nonumber\\
 &\leq& \frac{1}{2\tau}(1 + r) \sprod{u}{u}_H\; ,
\nonumber
\end{eqnarray}
where $r = \sqrt{\alpha\beta}$, and Proposition \ref{Prop:QuasiDiss} follows with $b = (1+r)/2$.

\subsection{Proof of Proposition \ref{Prop:DenseCond}}

Let $v = (\sigma,\bbB,\bbF,\bbV) \in
C^\infty(\bar{\Omega},\Real^{16})$ with $n^i B_i = 0$ on
$\partial\Omega$, and let $\lambda > 0$. We show that there exists $u
= (\rho,\bbA,\bbE,\bbW)\in D({\cal A})$ such that $(\lambda- {\cal A})u =
v$; that is, such that
\begin{eqnarray}
\lambda \rho &=& \sigma,
\label{Eq:lambdarho}\\
\lambda A_j &=& E_j - \nabla_j\Delta^{-1}\nabla^k E_k + B_j\; ,
\label{Eq:lambdaAj}\\
\lambda E_j &=& \nabla^i(W_{ij} - W_{ji}) + \alpha(\nabla_j W - \nabla^i W_{ji}) + F_j\; ,
\label{Eq:lambdaEj}\\
\lambda W_{ij} &=& \nabla_i E_j + \frac{\beta}{2}\, h_{ij} (\nabla^k E_k - \rho)
 - \nabla_i\nabla_j\Delta^{-1}\nabla^k E_k + V_{ij}\; .
\label{Eq:lambdaWij}
\end{eqnarray}
A consequence of this is the elliptic equation for $\bbE$
\begin{equation}
\lambda^2 E_j = \nabla^i\nabla_i E_j + (\alpha\beta-1)\nabla_j\nabla^i E_i + S_j\; ,
\end{equation}
where $S_j = \nabla^i(V_{ij} - V_{ji}) + \alpha(\nabla_j V - \nabla^i V_{ji}) + \lambda F_j - \alpha\beta\lambda^{-1}\nabla_j\sigma$,
with the boundary conditions on $\partial\Omega$
\begin{eqnarray}
&& \lambda H_i^{\;\; j} E_j + a_i^{\;\; j} n^k(\nabla_k E_j - \nabla_j E_k) 
 = - a_i^{\;\; j} n^k(V_{kj} - V_{jk}) .
\nonumber\\
&& \lambda\nabla^k E_k + b_0 r\, n^i\nabla_i\nabla^k E_k 
 = -b_0\frac{r}{\beta} n^i(\nabla_i V - \nabla^j V_{ij}) - b_0 r\lambda^{-1} n^i\nabla_i\sigma + \sigma.
\nonumber
\end{eqnarray}
In appendix \ref{App:Elliptic} it is proven in the case $a_i^{\;\; j}
= a_0 H_i^{\;\; j}$, where $a_0\in C^\infty(\partial\Omega)$ is a
strictly positive function, that for sufficiently large $\lambda > 0$
there exists a unique solution $\bbE\in C^\infty(\bar{\Omega})$ to
this problem. Setting $\rho = \lambda^{-1}\sigma$ and determining
$A_j$ and $W_{ij}$ from Eq. (\ref{Eq:lambdaAj}) and
Eq. (\ref{Eq:lambdaWij}), respectively, yields a solution $u\in
D({\cal A})$ to $(\lambda- {\cal A})u = v$. Since $\{
(\sigma,\bbB,\bbF,\bbV) \in C^\infty(\bar{\Omega},\Real^{16}) : n^i
B_i = 0 \hbox{ on $\partial\Omega$} \}$ is dense in $H$, proposition
\ref{Prop:DenseCond} follows.

\subsection{Intertwining operators and the constraint hypersurface}

Here we prove that the semigroup $P(t)$ constructed in theorem
\ref{Thm:Homo} leaves the constraint manifold, defined by
\begin{displaymath}
{\cal C} = \{ u = (\rho,A_i,E_i,W_{ij})\in H : W_{ij} = \nabla_i A_j\; , \nabla^i E_i = \rho \}
\end{displaymath}
invariant. In order to do so we introduce the Hilbert space $G = \{ v
= (C^{(W)}_{ij}, C, C_i)\in L^2(\Omega,\Real^{13}) \}$ and the linear
operator ${\cal B}: D({\cal B})\subset G \rightarrow G$ which is
defined by
\begin{equation}
D({\cal B}) = \{ (C^{(W)}_{ij}, C, C_k)\in C^\infty(\bar{\Omega},\Real^{13})
: \hbox{On $\partial\Omega$ we have } C - b_0 \frac{r}{\beta}\, n^i C_i = 0 \},
\label{Eq:DefDB}
\end{equation}
and
\begin{equation}
{\cal B}\left( \begin{array}{l} C^{(W)}_{ij} \\ C \\ C_k \end{array} \right) = 
\left( \begin{array}{l} -\frac{\beta}{2}\, h_{ij} C \\
 -\alpha \nabla^k C_k \\ -\beta\nabla_k C
 \end{array} \right).
\label{Eq:DefB}
\end{equation}
The operator ${\cal B}$ describes the evolution of the constraint
variables, cf.
Eqs. (\ref{Eq:Cijf},\ref{Eq:Cf},\ref{Eq:Ckf}). Similarly to before,
one shows that $D({\cal B})$ is dense in $G$, that ${\cal B}$ is
closable and that its closure generates a strongly continuous
semigroup $Q(t)$ in $G$. The Hilbert spaces $H$ and $G$ are related
by the linear operator $L: H \rightarrow G$ which is defined by
\begin{equation}
L\left( \begin{array}{l} \rho \\ A_j \\ E_j \\ W_{ij} \end{array} \right) = 
\left( \begin{array}{l} W_{ij} - \nabla_i A_j \\ \rho - \nabla^i E_i \\ 
\nabla_i W - \nabla^j W_{ij} 
 \end{array} \right).
\label{Eq:DefL}
\end{equation}
Since $L$ is continuous, the constraint hypersurface ${\cal C} = \{
u\in H : L(u) = 0 \}$ is a closed subset of $H$. Furthermore, we have
$L D({\cal A}) \subset D({\cal B})$, and we verify the intertwining
relation
\begin{equation}
L{\cal A} u = {\cal B} L u, \qquad
u\in D({\cal A}).
\end{equation}
Since $L$ is continuous and ${\cal A}$ and ${\cal B}$ are closable,
it also follows that $L D(\bar{\cal A}) \subset D(\bar{\cal B})$ and that
\begin{equation}
L\bar{\cal A} u = \bar{\cal B} L u, \qquad
u\in D(\bar{\cal A}).
\end{equation}
This implies that for each $u\in D(\bar{\cal A})$, the curve $v: \Real_+ \rightarrow D(\bar{\cal B}), t\mapsto L
P(t)u$ is differentiable and satisfies
\begin{displaymath}
\frac{d}{dt} v(t) = L\bar{\cal A} P(t) u = \bar{\cal B} L P(t) u = \bar{\cal B} v(t), \qquad
v(0) = Lu.
\end{displaymath}
Since $\bar{\cal B}$ is the generator of the semigroup $Q(t)$, it
follows that $L P(t) u = v(t) = Q(t) L u$, $t > 0$. Therefore,
\begin{equation}
L P(t) = Q(t) L, \qquad t \geq 0,
\label{Eq:Intertwining}
\end{equation}
since $D(\bar{\cal A})$ is dense in $H$. Eq. (\ref{Eq:Intertwining})
implies that $P(t)$ leaves the constraint manifold ${\cal C}$
invariant.

\subsection{Proof of Theorem \ref{Thm:Main}}

Finally, we discuss the case of nontrivial source functions and
boundary data. We first reduce the problem to the case of homogeneous
boundary conditions with nontrivial source functions and then use
Duhamel's principle to construct the solution of the resulting
problem.

Let $\bbJ(t)\in C^\infty(\bar{\Omega},\Real^3) \subset H_E$, $t > 0$,
be a continuous curve in $H_E$, let $u_0 =
(\rho^{(0)},\bbA^{(0)},\bbE^{(0)},\bbW^{(0)}) \in
C^\infty(\Omega,\Real^{16})\cap H$ be smooth initial data, and let
$g_i(t)\in C^\infty(\partial\Omega)$, $t > 0$, be a continuously
differentiable curve in $L^2(\partial\Omega)$ with the property that
$g_i(t) n^i = 0$ on $\partial\Omega$ for all $t > 0$. Assume that the
initial and boundary data satisfy the zeroth order compatibility
condition
\begin{eqnarray}
&& H_i^{\;\; j} E^{(0)}_j + a_i^{\;\; j} n^k(W^{(0)}_{kj} - W^{(0)}_{jk}) = g_j,
\label{Eq:CC1}\\
&& \nabla^k E^{(0)}_k - \rho^{(0)} + b_0 \frac{r}{\beta}\, n^i(\nabla_i W^{(0)} - \nabla^j W^{(0)}_{ij}) = 0,
\label{Eq:CC2}
\end{eqnarray}
on $\{ t=0 \} \times \partial\Omega$. Choose $\bbF(t)\in
C^\infty(\bar{\Omega},\Real^3)$ such that $H_i^{\;\; j} F_j(t) =
g_j(t)$ on $\partial\Omega$ and $\nabla^k F_k(t) = 0$ on $\Omega$ (see
Lemma \ref{Lem:SmallEBC}) for all $t > 0$. We can choose $\bbF(t)$
such that it describes a continuously differentiable curve in
$H^1(\Omega,\Real^3)$. Define $v(t) = (0,0,F_i(t),0)\in H$, ${\cal
S}(t) = (\nabla^i J_i(t),0,J_i(t),0)\in H$ and ${\cal F}(t) = {\cal
A}v(t) - \dot{v}(t) + {\cal S}(t)$, where ${\cal A}v(t)$ is defined by
Eq. (\ref{Eq:DefA}). One can verify that $v(t)$, ${\cal S}(t)$, ${\cal
F}(t)$, $t > 0$ define continuous curves in the Hilbert space
$H$. Next, consider the inhomogeneous Cauchy problem
\begin{eqnarray}
&& \frac{d}{dt} w = {\cal A} w + {\cal F},\\
&& w(0) = w_0 \equiv u_0 - v(0).
\end{eqnarray}
The compatibility conditions (\ref{Eq:CC1},\ref{Eq:CC2}) make sure
that $w_0 \in D({\cal A})$. We can solve this problem in an abstract
way using Duhamel's principle:
\begin{equation}
w(t) = P(t) w_0 + \int_0^t P(t-s) {\cal F}(s) ds,
\end{equation}
where $P(t)$ is the semigroup constructed in theorem \ref{Thm:Homo}
and where the above integral is a $H$-valued Riemann integral. By
construction, $u(t) = v(t) + w(t)$ satisfies the inhomogeneous
initial-boundary value problem (\ref{Eq:rhof},\ref{Eq:Af},\ref{Eq:Ef},\ref{Eq:Wf}),
(\ref{Eq:EllipticGauge}), (\ref{Eq:CPBC},\ref{Eq:PoyntingBC}) with
$u(0) = u_0$:
\begin{equation}
\frac{d}{dt} u = {\cal A} v - {\cal F} + {\cal S} + {\cal A} w + {\cal F}
 = {\cal A} u + {\cal S},
\end{equation}
and $u(t)$ satisfies the inhomogeneous boundary
conditions. Furthermore, using the fact that $L v(t) = 0$ and $L {\cal
S}(t) = 0$ for all $t\geq 0$, and using the intertwining relation
(\ref{Eq:Intertwining}), we find $L u(t) = Q(t) L u_0$, which shows
that the constraints are propagated. This concludes the proof of
Theorem \ref{Thm:Main}.

\section{Conclusions}
\label{sect:Conclusions}

In this article we considered the initial-boundary value problem for a
formulation of Maxwell's equations which is structurally similar to
the Einstein-Christoffel family of formulations of Einstein's field
equations. In particular, our problem includes constraints that
propagate with nontrivial speeds at timelike boundaries and so need to
be controlled by specifying constraint-preserving boundary conditions.
These conditions are supplemented with boundary conditions that
control the incoming electromagnetic radiation. We have first
considered the case of a fixed, a priori specified electromagnetic
potential $\phi$ and shown that the resulting initial-boundary value
problem is not well posed in the expected sense, even though the
Kreiss condition is verified. This is shown by an explicit
counterexample representing an electrostatic solution in a ``bad''
gauge. This gauge is ``bad'' in the sense that it gives rise to modes
growing linearly in $\omega t$, where $\omega$ are the frequency
components of the initial data. The counterexample also motivates how
to deal with these modes by choosing a different gauge which
eliminates them. In this new gauge, we are able to derive well posed
initial-boundary value formulations; that is, we show that a solution
exists in some appropriate Hilbert space, is unique, and depends
continuously on the initial and boundary data.

One key idea in our proof of well posedness is to start with an
estimate for the physical energy instead of estimating the norm
defined by the symmetrizer of the main evolution system, and with an
estimate that controls the constraint variables. This gives rise to
estimates for the gauge-invariant quantities. Once these quantities
are under control, one estimates the gauge-dependent quantities (in
our case the components of the vector potential and their first order
spatial derivatives). These a priori estimates then motivate the
Hilbert space in which the solutions are shown to exist using methods
from semigroup theory.

Unfortunately, the new gauge choice is of elliptic type, implying one
has to solve a scalar elliptic equation at each time step of a
numerical simulation. Although one can start an iterative method for
solving it with the prior time value as seed, the procedure has some
computational costs. Therefore, an interesting question is whether
there exists other gauge choices leading to a well posed
initial-boundary value problem which are easier to implement
numerically. The relevant two places where the gauge choice is
important is in estimating the $L^2$ norm of the vector potential
$A_i$ and of its derivatives $\nabla_i A_j$. For the first case, since
a $L^2$ estimate for $E_i$ is given from the estimates for the
gauge-invariant quantities, all what is needed is that the gradient of
$\phi$ be bounded in $L^2$ too. For the second case, the vanishing of
the normal component, $A_n$, of $A_i$ at the boundary is needed. In
particular, this is true if $A_n|_{\partial \Omega} = 0$ initially and
$n^k(\nabla_k \phi + E_k) |_{\partial \Omega} = 0$ for all times. So
any gauge fixing satisfying these conditions should lead to a well
posed system.

How much of the above discussion can be carried over to General
Relativity? As an example, consider the Einstein-Christoffel
formulations with an a priori given shift vector that is tangential to
the boundary. For this system, one has to specify six boundary
conditions. There are three conditions that are needed in order to
preserve the constraints \cite{CPRST}. Then, there are two conditions
that should control incoming gravitational radiation. At least in the
weak field approximation, where linearizations about Minkowski space
are considered, it should be possible to specify such conditions in
terms of the Weyl tensor, since -- in the weak field approximation --
this tensor is gauge invariant. Finally, there is a remaining boundary
condition that has to be used in order to control a gauge freedom. We
expect that for the a priori given shift case a similar counterexample
as the one we discussed here can be found and that the resulting
initial-boundary value formulation will not be well posed in the
expected sense. On the other hand, we also expect that this problem
can be avoided by requiring an appropriate elliptic equation for the
shift and that well posedness can be derived for the resulting
problem, at least for the case of linearizations about a stationary
background.

Finally, we briefly discuss a different toy model describing Maxwell's
equations which resembles more closely the BSSN-type of formulations
of General Relativity (see \cite{SN, BS} for the original references
and \cite{SCPT, NOR, GMG2, BeySar} for an analysis of the mathematical
structure of the equations). Instead of the variables $W_{ij}$ we
introduce an extra variable $\Gamma$ along with the constraint
$C^{\Gamma} \equiv \Gamma - \nabla^k A_k = 0$, and consider the mixed
first order-second order system
\begin{eqnarray}
\partial_t A_i &=& E_i + \nabla_i \phi, 
\label{Eq:Afs}\\ 
\partial_t E_j &=& \nabla^i\nabla_i A_j + \alpha \nabla_j\Gamma - (1+\alpha) \nabla_j\nabla^i A_i\; ,
\label{Eq:Efs}\\ 
\partial_t\Gamma &=& (1+\beta)\nabla^k E_k + \nabla^k\nabla_k\phi ,
\label{Eq:Gfs}
\end{eqnarray}
where $\alpha$, $\beta$ are parameters subject to the condition
$\alpha\cdot\beta > 0$. This model problem has been discussed by
several authors \cite{KWB, Fiske, GMG1, Calabrese} in the past. The
constraint variables $C = -\nabla^k E_k$, $C^{\Gamma}$ propagate
according to
\begin{eqnarray}
\partial_t C &=& -\alpha \nabla^i\nabla_i C^{\Gamma}, \\
\partial_t C^{\Gamma} &=& -\beta C .
\end{eqnarray}
In analogy to our previous model example, we consider the boundary
conditions
\begin{eqnarray}
&& V^{(+)} = c\; V^{(-)}, \qquad
   V^{(\pm)} \equiv C \mp \frac{\sqrt{\alpha\beta}}{\beta}\, n^k \nabla_k C^{\Gamma} \; ,
\label{Eq:CPBC_BSSN}\\
&& w_i^{(+)} = S_i^{\;\; j} w_j^{(-)}\; , \qquad
   w_i^{(\pm)} \equiv H_i^{\;\; j} E_j \pm n^k (\nabla_k A_i - \nabla_i A_k),
\label{Eq:PoyntingBC_BSSN}
\end{eqnarray}
where $|c|\leq 1$, and $S_i^{\;\; j}$ is a smooth matrix-valued
function on $\partial\Omega$ with the properties that $S_i^{\;\; j}
n^i = 0$, $S_i^{\;\; j} n_j = 0$ and $H^{ij} S_i^{\;\; r} S_j^{\;\; s}
\leq H^{rs}$. There is an analogous example to the one presented in
section \ref{sect:MP} which shows that it is not possible to prove
well posedness in a space that controls the $L^2$ norm of the fields
$A_i$, $E_i$, $\Gamma$ and the first spatial derivatives of $A_i$ when
the gauge $\phi=0$ is adopted. However, it is not difficult to see
that one can adapt the estimates of section \ref{sect:EE} when the
elliptic gauge condition (\ref{Eq:EllipticGauge}) is imposed instead,
and so one can proceed as in this article to show well posedness.

\section{Acknowledgements}

It is a pleasure to thank G. Calabrese, L. Lehner, L. Lindblom,
G. Nagy, M. Scheel, M. Tiglio and M. Tom for useful comments and
discussions. OS particularly thanks H. Beyer for many illuminating
discussions on the use of semigroup methods and for communicating his
manuscript \cite{BC-preprint}. We also thank the Caltech Visitors
Program for the Numerical Simulation of Gravitational Wave Sources,
the Klavi Institute for Theoretical Physics Visitor Program:
Gravitational Interaction of Compact Objects, and the Isaac Newton
Institute for Mathematical Sciences, Cambridge Visitor Program:
Hyperbolic Models in Astrophysics and Cosmology for financial
support. This work was supported in part by the Center for Computation
\& Technology at Louisiana State University, by grants NSF-PHY0244335,
NSF-PHY0244699, NSF-PHY0312049, NSF-PHY-0326311 NSF-PHY0099568,
NSF-INT-0204937, NSF-INT-0307290, NASA-NAG5-13430, by funds from the
Horace Hearne Jr. Laboratory for Theoretical Physics, and by CONICET;
SECYT-UNC; and Agencia C\'ordoba Ciencia.

\appendix
\section{Elliptic problem}
\label{App:Elliptic}

Here we show existence of smooth solutions to the elliptic boundary
value problem which arises in the proof of Proposition
\ref{Prop:DenseCond}. We first simplify the problem and then apply
standard methods to analyze the resulting boundary value problem.

Let $\Omega$ be a bounded open subset of $\Real^3$ with $C^\infty$
boundary $\partial\Omega$. In the following, $\bbn = n_i dx^i$ denotes
the outward unit co-normal to $\partial\Omega$, and $\nabla$ denotes
the covariant derivative with respect to the Eulerian metric $ds^2 =
\delta_{ij} dx^i dx^j$ on $\Real^3$. We also introduce the projection
operator $H_i^j = \delta_i^j - n_i n^j$ on the tangent space of
$\partial\Omega$, and the extrinsic curvature of the boundary surface,
$\kappa_{ij} = H_i^r H_j^s \nabla_r n_s$.

Let $\lambda$ and $r$ be strictly positive constants, and let $c_0$
and ${\bf d}$ be a positive $C^\infty$ function and a matrix-valued
$C^\infty$ function, respectively, defined on the boundary of
$\Omega$, where ${\bf d} = (d_j^{\;\; k})$ satisfies $n^j d_j^{\;\; k}
= 0$, $d_j^{\;\; k} n_k = 0$. We are interested in the following
boundary value problem:
\begin{eqnarray}
\lambda^2 E_j - \nabla^i\nabla_i E_j - (r^2 - 1)\nabla_j\nabla^i E_i = S_j
&& \hbox{on $\Omega$},
\label{Eq:Elliptic}\\
(n^j\nabla_j + \lambda\, c_0)\nabla^i E_i = G && \hbox{on $\partial\Omega$},
\label{Eq:BC1}\\
n^i(\nabla_i E_j - \nabla_j E_i) + \lambda\, d_j^{\;\; i} E_k = g_k && \hbox{on $\partial\Omega$},
\label{Eq:BC2}
\end{eqnarray}
where $\bbS\in C^\infty(\bar{\Omega},\Real^3)$ and $G\in
C^\infty(\partial\Omega,\Real)$, $\bbg\in
C^\infty(\partial\Omega,\Real^3)$ are given functions with the
property that $n^j g_j = 0$. Notice that this has the form of the
problem arising in Proposition \ref{Prop:DenseCond} with $r =
\sqrt{\alpha\beta}$,
\begin{eqnarray}
c_0 &=& \frac{1}{r b_0} = \frac{1}{r} \frac{1-c}{1+c}\; ,
\nonumber\\
{\bf d} &=& {\bf a}^{-1} = {\bf H}({\bf H} + {\bf S})^{-1} ({\bf H} - {\bf S}).
\nonumber
\end{eqnarray}
In this appendix, we prove

\begin{theorem}
\label{Thm:MainElliptic}
If $c_0\geq 0$ and ${\bf d} = d_0 {\bf H}$ where $d_0$ is a strictly
positive function on $\partial\Omega$, and if $\lambda > 0$ is
sufficiently large, the boundary value problem (\ref{Eq:Elliptic}),
(\ref{Eq:BC1}), (\ref{Eq:BC2}) possesses a unique solution $\bbE \in
C^\infty(\bar{\Omega},\Real^3)$.
\end{theorem}

We first simplify the problem and show that it is, basically,
sufficient to consider the case where $r=1$ and $c_0$ is replaced by
an arbitrary positive $C^\infty$ function $\tilde{c}_0$ on
$\partial\Omega$. This freedom in the choice of $\tilde{c}_0$ will be
important in view of constructing coercive quadratic forms.

Let $\bbS\in C^\infty(\bar{\Omega},\Real^3)$ and $G\in
C^\infty(\partial\Omega,\Real)$, $\bbg\in
C^\infty(\partial\Omega,\Real^3)$ be given. A solution to the problem
(\ref{Eq:Elliptic}), (\ref{Eq:BC1}), (\ref{Eq:BC2}) can be obtained by
performing the following steps:
\begin{enumerate}
\item
Decompose $\bbS = \bbS^T + \bbnab J$ where $\bbS^T\in
C^\infty(\bar{\Omega},\Real^3)$, $J \in C^\infty(\bar{\Omega},\Real)$
satisfy $\dvrg\bbS^T \equiv \nabla^i S^T_i = 0$ and $\left. J
\right|_{\partial\Omega} = 0$ (this requires solving the Dirichlet
problem $\Delta J = \dvrg\bbS$, $\left. J \right|_{\partial\Omega} =
0$).

\item
Define $\tilde{G} = \lambda^{-2}\left. (r^2 G + n^i\nabla_i J)
\right|_{\partial\Omega}$ and solve the problem
\begin{eqnarray}
(\lambda^2 - r^2\nabla^i\nabla_i)\phi = J 
&& \hbox{on $\Omega$},
\label{Eq:Boxphi}\\
(n^j\nabla_j + \lambda\, c_0)\phi = \tilde{G} && \hbox{on $\partial\Omega$},
\label{Eq:BCphi}
\end{eqnarray}
Using similar techniques as the one used below, it is not difficult to
show that this problem has a unique solution $\phi\in
C^\infty(\bar{\Omega},\Real)$ if $r > 0$, $c_0 \geq 0$.

\item
Solve the boundary value problem
\begin{eqnarray}
\lambda^2 F_j - \nabla^i\nabla_i F_j = S^T_j 
&& \hbox{on $\Omega$},
\label{Eq:BoxF}\\
(n^j\nabla_j + \lambda\,\tilde{c}_0)\nabla^i F_i = 0 && \hbox{on $\partial\Omega$},
\label{Eq:BC1F}\\
n^i(\nabla_i F_j - \nabla_j F_i) + \lambda\, d_j^{\;\; k} F_k 
 = g_j - \lambda\, d_j^{\;\; k}\nabla_k\phi
 && \hbox{on $\partial\Omega$},
\label{Eq:BC2F}
\end{eqnarray}
where $\tilde{c}_0$ is an arbitrary positive and smooth function on
$\partial\Omega$.  Notice that a solution $\bbF\in
C^\infty(\bar{\Omega},\Real^3)$ satisfies $(\lambda^2 -
\nabla^i\nabla_i)\dvrg\bbF = 0$, $\left. (n^j\nabla_j +
\lambda\,\tilde{c}_0)\dvrg\bbF = 0 \right|_{\partial\Omega}$, so
$\dvrg\bbF = 0$ according to the previous step.
\item
Then, it is easy to verify that $\bbE = \bbF + \bbnab\phi$ solves the
boundary-value problem (\ref{Eq:Elliptic}), (\ref{Eq:BC1}),
(\ref{Eq:BC2}). This solution is unique, because given two solutions
their difference $\bbE^\Delta$ satisfies $(\lambda^2 -
r^2\nabla^i\nabla_i)\dvrg\bbE^\Delta = 0$, $\left. (n^j\nabla_j +
\lambda\, c_0)\dvrg\bbE^\Delta = 0 \right|_{\partial\Omega}$, so
$\dvrg\bbE^\Delta = 0$ by the second step, and so $\bbE^\Delta$
satisfies the homogeneous version of the problem in the third step for
which uniqueness will be shown.

\end{enumerate}
Therefore, it is sufficient to prove the statement of theorem
\ref{Thm:MainElliptic} for the case where $r=1$ and where $c_0$ is one
particular positive and smooth function. Before we attack the problem,
it turns out to be convenient to rewrite the boundary condition
(\ref{Eq:BC1}) in a different form. Using Eq. (\ref{Eq:Elliptic}) with
$r=1$, we rewrite
\begin{eqnarray}
n^j\nabla_j \nabla^i E_i &=& n^j(n^i n^k + H^{ik})\nabla_j\nabla_i E_k
\nonumber\\
 &=& (\lambda^2 E_n - S_n) + H^{ik}\nabla_i(n^j\nabla_j E_k - n^j\nabla_k E_j) 
 + \kappa^{ik}(\nabla_i E_k - \nabla_k E_i)
\nonumber\\
 &=& (\lambda^2 E_n - S_n) + H^{ik}\nabla_i( g_k - \lambda\, d_k^{\;\; l} E^{tg}_l),
\nonumber
\end{eqnarray}
where $E_n = E_i n^i$, $E^{tg}_l = H_l^k E_k$, and where we have used
the fact that $\kappa_{ij}$ is symmetric. Combining this with
$\nabla^j E_j = n^i(n^j\nabla_j + \kappa) E_i + H^{ik}\nabla_i
E^{tg}_k$ the boundary condition (\ref{Eq:BC1}) can be rewritten in
the form
\begin{displaymath}
n^i( n^j\nabla_j + \kappa + \lambda c_0^{-1}) E_i
 + ( H^{ik} - c_0^{-1} d^{ik} ) D_i E^{tg}_k 
 - c_0^{-1} (D_i d^{ik}) E^{tg}_k = G_0\; ,
\end{displaymath}
where $G_0 = (\lambda\, c_0)^{-1}(G + S_n - D^i g_i)$ and where $D$
denotes the covariant derivative with respect to the induced metric
$H_{ij}$ on $\partial\Omega$.  Finally, we combine this with the
boundary condition (\ref{Eq:BC2}) and obtain the following boundary
value problem:
\begin{eqnarray}
\lambda^2 E_j - \nabla^i\nabla_i E_j = S_j
&& \hbox{on $\Omega$}, \quad
\label{Eq:BoxE}\\
n^i\nabla_i E_j - D_j E_n + n_j( H^{ik} - c_0^{-1} d^{ik})D_i E_k^{tg} 
 + \mu_j^{\;\; k} E_k^{tg} + \eta\, n_j E_n - c_0^{-1} n_j (D_i d^{ik}) E^{tg}_k
 = n_j G_0 + g_j
&& \hbox{on $\partial\Omega$}, \quad
\label{Eq:BCE}
\end{eqnarray}
where $\mu_{jk} = \lambda\, d_{jk} + \kappa_{jk}$ and $\eta = \lambda
c_0^{-1} + \kappa$. 

From now on, we specialize to the case where $d_{kj} = d_0 H_{kj}$
where $d_0$ is a strictly positive and smooth function on
$\partial\Omega$. Although we believe that this restriction is not
necessary, it simplifies the proof below. Choose $c_0 = d_0/2$ in
which case the boundary condition (\ref{Eq:BCE}) reduces to
\begin{equation}
n^i\nabla_i E_j - D_j E_n - n_j D^s E_s^{tg} 
 + \mu_j^{\;\; k} E_k^{tg} + \eta\, n_j E_n - 2 n_j d_0^{-1}(D^i d_0) E^{tg}_i
 = n_j G_0 + g_j
\qquad \hbox{on $\partial\Omega$},
\label{Eq:BCEbis}
\end{equation}
where $\mu_{jk} = \lambda\, d_0 H_{jk} + \kappa_{jk}$ and $\eta =
2\lambda d_0^{-1} + \kappa$. Notice that the third term in the
boundary condition (\ref{Eq:BCEbis}) is minus the formal adjoint with
respect to $L^2(\partial\Omega)$ of the second term in
(\ref{Eq:BCEbis}). This and the positivity of $\mu_{ij}$ and $\eta$
will allow us to find the appropriate elliptic estimates and to prove

\begin{theorem}
\label{Thm:Simpl}
Let $d_0\in C^\infty(\partial\Omega)$ be strictly positive at each
point of $\partial\Omega$. Denote by $\kappa_1 \leq \kappa_2$ the
nonzero eigenvalues of $\kappa_{ij}$ at each point of
$\partial\Omega$, and set $N = \sqrt{ H^{ij} D_i\log d_0 \cdot D_j\log
d_0 }$. Then, if $\lambda$ is sufficiently large such that
\begin{equation}
\lambda \geq \max\{ 0, \frac{1}{2}\left( N - (\kappa_1+\kappa_2)d_0 \right), 
  N - \frac{\kappa_1}{d_0} \} 
\end{equation}
the boundary value problem (\ref{Eq:BoxE}), (\ref{Eq:BCEbis})
possesses a unique solution $\bbE \in C^\infty(\bar{\Omega},\Real^3)$.
\end{theorem}

The proof uses standard methods \cite{Treves} \cite{Taylor} \cite{RR}:
We first recast the boundary value problem in weak form and show the
existence and uniqueness of weak solutions. After that, we derive
regularity results which show that this solution is, in fact, smooth.

\subsection{Weak formulation}

In the following, let $H^k(\Omega,\Real^m)$, $k\geq 0$, denote the
Sobolev space of functions $\Omega \rightarrow \Real^m$ whose partial
derivatives of order $\leq k$ are in $L^2(\Omega)$.  We abbreviate
$W^k = H^k(\Omega,\Real^3)$. Introduce the linear operators $L: W^{k+2} \rightarrow W^k$,
$b: W^{k+2} \rightarrow H^{k+1/2}(\partial\Omega,\Real^3)$ defined by
\begin{eqnarray}
L(\bbE)_j &=& (\lambda^2 - \nabla^i\nabla_i) E_j, 
\nonumber\\
b(\bbE)_j &=& n^i\nabla_i E_j - D_j E_n - n_j D^s E_s^{tg} 
 + \mu_j^{\;\; k} E_k^{tg} + \eta\, n_j E_n - 2 n_j d_0^{-1}(D^i d_0) E^{tg}_i\, ,
\nonumber
\end{eqnarray}
We consider the boundary value problem $L(\bbE)_j = S_j$, $b(\bbE)_j =
g_j$, where $\bbS\in W^k$ and $\bbg\in
H^{k+1/2}(\partial\Omega,\Real^3)$.

Let $\bbE, \bbF \in C^\infty(\bar{\Omega},\Real^3)$. Using Gauss'
theorem we find
\begin{eqnarray}
\int_\Omega F^j L(\bbE)_j d^3 x &=& \int_\Omega \left( \lambda^2 F^j E_j + \nabla^i F^j\cdot \nabla_i E_j \right) d^3 x
 - \int_{\partial\Omega} F^j n^i\nabla_i E_j dS
\nonumber\\
 &=& Q_\lambda(\bbE,\bbF) - \int_{\partial\Omega} F^j b(\bbE)_j \, dS
\nonumber
\end{eqnarray}
where $dS$ denotes the surface element on $\partial\Omega$ and where
we have defined the bilinear form
\begin{eqnarray}
Q_\lambda(\bbE,\bbF) &=& \int_\Omega \left( \lambda^2 E^j F_j + \nabla^i E^j\cdot \nabla_i F_j \right) d^3 x
\nonumber\\
 &+& \int_{\partial\Omega} \left[ \mu^{ij} E_i F_j + \eta\, E_n F_n
 - 2 F_n (D^i\log d_0) E^{tg}_i
 - F_n D^i E^{tg}_i - F^{tg}_i D^i E_n \right] dS.
\nonumber
\end{eqnarray}
Therefore,
\begin{equation}
(\bbF, L(\bbE) )_{L^2(\Omega)} + (\bbF, b(\bbE) )_{L^2(\partial\Omega)} = Q_\lambda(\bbE,\bbF).
\label{Eq:WeakPreForm}
\end{equation}
Because of the trace theorems we can extend $Q_\lambda$ to $W^1 \times
W^1$: If $\bbE, \bbF\in W^1$, they are also in
$H^{1/2}(\partial\Omega)$, and (in the sense of distributions) $D_i
E_n \in H^{-1/2}(\partial\Omega)$, which is the dual space of
$H^{1/2}(\partial\Omega)$, so the integral $\int_{\partial\Omega}
F^{tg}_i D^i E_n dS$ makes sense. Furthermore, there is a constant $C$
such that
\begin{equation}
| Q_\lambda(\bbE,\bbF) | \leq C \| \bbE \|_{W^1} \cdot \| \bbF \|_{W^1}\;  \hbox{for all $\bbE, \bbF \in W^1$}.
\end{equation}
Thus $Q_\lambda$ is a bounded bilinear form on the Hilbert space
$W^1$. Equation (\ref{Eq:WeakPreForm}), which holds for all $\bbE\in
W^2$, $\bbF\in W^1$, implies

\begin{lemma}
\label{Lem:WeakStrong}
Let $\bbE\in W^2$, and suppose that $\bbS\in W^0$, $\bbg\in
H^{1/2}(\partial\Omega,\Real^3)$ are given. Then, $\bbE$ is a (strong)
solution of (\ref{Eq:BoxE}), (\ref{Eq:BCEbis}) if and only if
\begin{equation}
(\bbF, \bbS )_{L^2(\Omega)} + (\bbF, \bbg )_{L^2(\partial\Omega)} = Q_\lambda(\bbE,\bbF).
\label{Eq:WeakForm}
\end{equation}
for all $\bbF\in W^1$.
\end{lemma}

{\bf Proof}: 
The ``only if'' part is clear. In order to show the ``if'' part assume
that (\ref{Eq:WeakForm}) holds for all $\bbF\in W^1$. In particular,
it holds for all $\bbF\in C_0^\infty(\bar{\Omega},\Real^3)$.  Since
$C_0^\infty(\bar{\Omega},\Real^3)$ is dense in $L^2(\Omega,\Real^3)$
it follows, in view of Eq. (\ref{Eq:WeakPreForm}) that $L(\bbE) =
\bbS$. Next, choose $\bbF\in C^\infty(\bar{\Omega},\Real^3)$ with $F_j
\neq 0$ at the boundary, which implies $b(\bbE)_j = g_j$.
\qed

This motivates the following

\begin{definition}
Suppose $\lambda > 0$, and let $\bbS\in W^0$, $\bbg\in
H^{1/2}(\partial\Omega,\Real^3)$ be given. We call $\bbE\in W^1$ a
{\em weak solution} of (\ref{Eq:BoxE}), (\ref{Eq:BCEbis}) if equation
(\ref{Eq:WeakForm}) holds for all $\bbF\in W^1$.
\end{definition}

The next lemma implies the existence of weak solutions:
\begin{lemma}
\label{Lem:Coercive}
Suppose the assumptions of Theorem \ref{Thm:Simpl} are met. Then,
$Q_\lambda$ is coercive. That is, there exists a constant $a_0 =
a_0(\lambda) > 0$ such that
\begin{equation}
Q_\lambda(\bbE,\bbE) \geq a_0 \| \bbE \|_{W^1}^2
\end{equation}
for all $\bbE\in W^1$.
\end{lemma}

{\bf Proof}:
Let $\bbE\in C^\infty(\bar{\Omega},\Real^3)$. We have
\begin{eqnarray}
Q_\lambda(\bbE,\bbE) &=& \int_\Omega \left( \lambda^2 E^j E_j + \nabla^i E^j\cdot \nabla_i E_j \right) d^3 x
\nonumber\\
 &+& \int_{\partial\Omega} \left[ \mu^{ij} E_i E_j + \eta\, E_n^2
 - 2 E_n (D^i\log d_0) E^{tg}_i
 - E_n D^i E^{tg}_i - E^{tg}_i D^i E_n \right] dS.
\nonumber
\end{eqnarray}
The last two terms in the boundary integral form a total divergence,
and so their contribution to the integral vanish. Therefore, setting
$a_0 = \min\{ \lambda^2, 1 \}$, $N = \sqrt{ H^{ij} D_i\log d_0 \cdot
D_j\log d_0 }$ and using Schwarz' inequality,
\begin{eqnarray}
Q_\lambda(\bbE,\bbE) &\geq& a_0 \| \bbE \|_{W^1}^2
 + \int_{\partial\Omega} \left( \mu^{ij} E_i E_j + \eta\, E_n^2 - N \left[ d_0 H^{ij} E_i E_j + d_0^{-1} E_n^2 \right] \right) dS
\nonumber\\
 &=& a_0 \| \bbE \|_{W^1}^2
 + \int_{\partial\Omega} \left( \left[ (\lambda-N)d_0 H^{ij} + \kappa^{ij} \right] E_i E_j 
  + \left[ (2\lambda-N)d_0^{-1} + \kappa \right] E_n^2 \right) dS.
\nonumber\\
 &\geq& a_0 \| \bbE \|_{W^1}^2.
\nonumber
\end{eqnarray}
Since $C^\infty(\bar{\Omega},\Real^3)$ is dense in $W^1$ and since
$Q_\lambda$ is bounded, the lemma follows.  
\qed

\begin{corollary}[Existence and Uniqueness of weak solutions]
Suppose the assumptions of Theorem \ref{Thm:Simpl} are met. Then, the
problem (\ref{Eq:BoxE}), (\ref{Eq:BCEbis}) possesses a unique weak
solution.
\end{corollary}

{\bf Proof}:
Define the linear functional $J\in W^{1*}$ by
\begin{displaymath}
J(\bbF) = (\bbF, \bbS )_{L^2(\Omega)} + (\bbF, \bbg
)_{L^2(\partial\Omega)}\; ,\qquad \bbF\in W^1.
\end{displaymath}
According to the Lax-Milgram lemma there is a unique $\bbE$ such that
$Q_\lambda(\bbE,.) = J$.
\qed

For later use, we introduce the operator ${\cal L}_\lambda: W^1
\rightarrow W^{1*}$ which is defined by
\begin{displaymath}
({\cal L}_\lambda\bbE)(\bbF) = Q_\lambda(\bbE,\bbF)
\end{displaymath}
for $\bbE$, $\bbF\in W^1$. Since $| Q_\lambda(\bbE,\bbF) | \leq C \|
\bbE \|_{W^1} \cdot \| \bbF \|_{W^1}$ it follows that $\| {\cal
L}_\lambda\bbE \|_{W^{1*}} \leq C \| \bbE \|_{W^1}$, so ${\cal
L}_\lambda$ is bounded. Furthermore, it is injective because of
$({\cal L}_\lambda\bbE)(\bbE) = Q_\lambda(\bbE,\bbE) \geq a_0 \| \bbE
\|_{W^1}^2$ and onto because of the implications of the Lax-Milgram
lemma. The weak form of (\ref{Eq:BoxE}), (\ref{Eq:BCEbis}) is simply
${\cal L}_\lambda\bbE = J$.

The following estimate will be important:

\begin{lemma}
${\cal L}_\lambda: W^1 \rightarrow W^{1*}$ defines a linear bounded
and bijective operator. For each $\bbE\in W^1$ it satisfies
\begin{equation}
\| \bbE \|_{W^1} \leq a_0^{-1} \| {\cal L}_\lambda\bbE \|_{W^{1*}}\; .
\label{Eq:W1*Estimate}
\end{equation}
\end{lemma}

{\bf Proof}:
According to Lemma \ref{Lem:Coercive} we have, for $\bbE\in W^1$,
\begin{displaymath}
a_0 \| \bbE \|_{W^1}^2 \leq Q_\lambda(\bbE,\bbE) = ({\cal L}_\lambda\bbE)(\bbE)
 \leq \| {\cal L}_\lambda\bbE \|_{W^{1*}} \cdot \| \bbE \|_{W^1}\; .
\end{displaymath}
\qed

\subsection{Regularity}

\begin{theorem}[Regularity]
\label{Thm:Regularity}
Suppose the assumptions of Theorem \ref{Thm:Simpl} are satisfied. Let
$k\geq 0$ and suppose that $\bbS\in W^k$, $\bbg\in
H^{k+1/2}(\partial\Omega,\Real^3)$.  Then, the unique weak solution
$\bbE$ of the boundary value problem (\ref{Eq:BoxE}), (\ref{Eq:BCEbis})
lies in $W^{k+2}$.  Furthermore, there is a constant $C_k > 0$ such
that
\begin{equation}
\| \bbE \|_{W^{k+2}} \leq C_k\left( \| \bbS \|_{W^k}  + \| \bbg \|_{H^{k+1/2}(\partial\Omega)} 
 + \| \bbE \|_{W^{k+1}} \right).
\label{Eq:WkEstimate}
\end{equation}
\end{theorem}

{\bf Remark}: 
For $k=0$ in particular, the theorem shows that any weak solution lies
in $W^2$ and so is a strong solution according to Lemma
\ref{Lem:WeakStrong}.\\

{\bf Remark}: 
Since the solution is unique, one can show (see, for example, Lemma
8.38 in Ref. \cite{RR}) that the last term in the right-hand side of
the estimate (\ref{Eq:WkEstimate}) can be dropped.  The resulting
estimate then implies well posedness in the sense that the solution
depends continuously on the data.\\

We break the proof of the theorem into several steps. First, we choose
a finite covering of $\Omega$ and show that the problem can be
localized. Applying difference quotients to the estimates
(\ref{Eq:W1*Estimate}), (\ref{Eq:WkEstimate}) we then show interior
regularity first. Next, we flatten the boundary by choosing an
appropriate coordinate patch and map the problem to a half plane
problem. Using difference quotients again, we first show
differentiability in the directions tangential to the boundary, and
then use the solution properties to show differentiability in the
normal directions as well.

We start with the case $k=0$.

\subsection{Localization}

Suppose $\bbE\in W^1$ is a weak solution, and let $\chi$ be the
restriction on $\Omega$ of a function in $C_0^\infty(\Real^3)$. Then
$\chi\cdot\bbF\in W^1$ for each $\bbF\in W^1$ and $J(\chi\bbF) =
({\cal L}_\lambda\bbE)(\chi\bbF) = ({\cal L}_\lambda(\chi\bbE))(\bbF)
+ R(\bbE,\bbF)$, where, using the Leibnitz rule and Gauss' theorem,
one finds
\begin{eqnarray}
R(\bbE,\bbF) &=& Q_\lambda(\bbE,\chi\bbF) - Q_\lambda(\chi\bbE,\bbF)
\nonumber\\
 &=& \int_{\Omega} \left[ \nabla^i E^j\cdot\nabla_i\chi\cdot F_j + \nabla_i(\nabla^i\chi\cdot E^j)\cdot F_j \right] d^3 x
\nonumber\\
 &+& \int_{\partial\Omega} \left[ D^s\chi( F_n E_s^{tg} + E_n F_s^{tg} ) - \nabla_n\chi\cdot E^j F_j \right] dS.
\nonumber
\end{eqnarray}
Therefore, ${\cal L}_\lambda(\chi\bbE) = \tilde{J}$ where
\begin{displaymath}
\tilde{J}(\bbF) = (\bbF, \tilde{\bbS} )_{L^2(\Omega)} + (\bbF, \tilde{\bbg})_{L^2(\partial\Omega)} 
\end{displaymath}
with
\begin{eqnarray}
\tilde{S}_j &=& \chi S_j - 2\nabla_i\chi\cdot\nabla^i E_j - (\nabla^i\nabla_i\chi) E_j \; ,
\label{Eq:Stilde}\\
\tilde{g}_j &=& \chi g_j 
 + \left[ \nabla_n\chi\cdot E_j - D_j\chi\cdot E_n - n_j D^s\chi\cdot E_s^{tg} \right]_{\partial\Omega}\; ,
\label{Eq:gtilde}
\end{eqnarray}
Since $\bbE\in W^1$, we have $\tilde{\bbS}\in W^0$ and
$\tilde{\bbg}\in H^{1/2}(\partial\Omega,\Real^3)$, and $\tilde{\bbS}$,
$\tilde{\bbg}$ have the same support as $\chi$. Therefore, it is
sufficient to assume that $\bbE$ and the data $\bbF$, $\bbg$ are
supported in some coordinate patch.

Finally, we notice that taking $\bbF\in C_0^\infty(\Omega,\Real^3)$ in
equation (\ref{Eq:WeakForm}) implies that
\begin{equation}
(\lambda^2 - \nabla^i\nabla_i) E_j = S_j
\label{Eq:WeakSol}
\end{equation}
in the sense of distributions.

\subsection{Interior regularity}

Assume first that $\bbE$ and $\bbS$ are supported in $\Omega$ (and so
the boundary data vanishes). Then, we can extend them to elements in
$H^1(\Real^3,\Real^3)$, and $H^0(\Real^3,\Real^3)$, respectively. By
rewriting $(\lambda^2 - \nabla^i\nabla_i) E_j = S_j$ in the form $(1 -
\nabla^i\nabla_i)E_j = S_j + (1 - \lambda^2)E_j$, it follows
immediately from the next lemma that $\bbE\in H^2(\Real^3,\Real^3)$.

\begin{lemma}
For each $s\in\Real$, the operator $1-\nabla^i\nabla_i$ induces an isometry
of $H^s(\Real^3)$ into $H^{s-2}(\Real^3)$.
\end{lemma}

{\bf Proof}: This follows immediately using Fourier transformation.
\qed

In view of the generalization to operators with variable coefficient
and in preparation of the boundary regularity result, we give a
different proof which is based on the use of difference quotients. So
let $\bbE\in H^1(\Real^3,\Real^3)$ be a weak solution with compact
support. The idea is to apply the estimate (\ref{Eq:W1*Estimate}) to
the difference quotient $D_i^h\bbE$, defined by
\begin{displaymath}
D_i^h\bbE(x) = \frac{1}{h}\left( \bbE(x + h e_i) - \bbE(x) \right),
\end{displaymath} 
where $h > 0$ and $e_i$ denote the unit standard vectors in $\Real^3$.
The partial derivative $D_i\bbE$ exists in the weak sense if and only
if $D_i^h\bbE$ is bounded in $h$ \cite{RR}. According to
Eq. (\ref{Eq:W1*Estimate}), we have
\begin{equation}
\| D_i^h\bbE \|_{H^1} \leq a_0^{-1} \| {\cal L}_\lambda D_i^h\bbE \|_{H^{1*}}\; .
\end{equation}
Now for any $\bbF\in H^1(\Real^3,\Real^3)$,
\begin{displaymath}
({\cal L}_\lambda D_i^h\bbE)(\bbF) = Q_\lambda(D_i^h\bbE,\bbF) = Q_\lambda(\bbE, D_i^{-h}\bbF) + \tilde{R}(\bbE,\bbF),
\end{displaymath}
where
\begin{displaymath}
\tilde{R}(\bbE,\bbF) = Q_\lambda(D_i^h\bbE,\bbF) - Q_\lambda(\bbE, D_i^{-h}\bbF).
\end{displaymath}
Using the ``Leibnitz rule'' and ``integration by parts'' for
differences quotients \cite{RR} it is not difficult to see that there
is a constant $C > 0$ (which depends only on the metric coefficients
and their derivatives) such that $|\tilde{R}(\bbE,\bbF)| \leq C \|
\bbE \|_{H^1}\cdot \| \bbF \|_{H^1}$. Furthermore, since $\bbE$ is a
weak solution,
\begin{eqnarray}
Q_\lambda(\bbE, D_i^{-h}\bbF) &=& J(D_i^{-h}\bbF)
\nonumber\\
 &\leq& \| \bbS \|_{L^2(\Real^3)} \cdot \| D_i^{-h}\bbF \|_{L^2(\Real^3)}
\nonumber\\
 &\leq& \| \bbS \|_{L^2(\Real^3)} \cdot \| \bbF \|_{H^1(\Real^3)}\; .
\label{Eq:J}
\end{eqnarray}
Taking everything together, we obtain
\begin{equation}
\| D_i^h\bbE \|_{H^1(\Real^3)} \leq a_0^{-1}\left( \| \bbS \|_{L^2(\Real^3)} + C \| \bbE \|_{H^1(\Real^3)} \right).
\end{equation}
Taking the limit $h \rightarrow 0$ we obtain $D_i\bbE\in
H^1(\Real^3,\Real^3)$.  Repeating the argument for each $i=1,2,3$, it
follows that $\bbE\in H^2(\Real^3,\Real^3)$ and that
\begin{equation}
\| \bbE \|_{H^2(\Real^3)} \leq C_1\left( \| \bbS \|_{L^2(\Real^3)} + \| \bbE \|_{H^1(\Real^3)} \right),
\end{equation}
for some constant $C_1$.

\subsection{Boundary regularity}

Let $\bbx_0\in \partial\Omega$ and suppose that $\bbE$ and the data
$\bbS$, $\bbg$ vanish for all $|\bbx - \bbx_0| > \delta > 0$.  If
$\delta$ is sufficiently small, we can introduce new coordinates
$\bby$ such that the boundary is described by $\bby = (y^1,y^2,y^3) =
(0,y^2,y^3)$. That is, we can consider the problem on the half plane
$\Real^3_+$. In order to show regularity, we proceed as above, except
that now we are only allowed to take the difference quotients
$D_A^i\bbE$, $A = 2,3$ in the directions tangent to the boundary.
Also, instead of Eq. (\ref{Eq:J}), we have
\begin{eqnarray}
Q_\lambda(\bbE, D_A^{-h}\bbF) &=& J(D_A^{-h}\bbF)
\nonumber\\
 &\leq& \| \bbS \|_{L^2(\Real^3_+)} \cdot \| D_A^{-h}\bbF \|_{L^2(\Real^3_+)} 
 + \| \bbg \|_{H^{1/2}(\Real^2)} \cdot \| D_A^{-h}\bbF \|_{H^{-1/2}(\Real^2)}
\nonumber\\
 &\leq& \left( \| \bbS \|_{L^2(\Real^3_+)} + \| \bbg \|_{H^{1/2}(\Real^2)} \right) \| \bbF \|_{H^1(\Real^3_+)}\; ,
\nonumber
\end{eqnarray}
where we have used the trace theorem in the last step. Therefore, it
follows that $D_A\bbE\in H^1(\Real^3_+,\Real^3)$, $A=2,3$. In order to
show that $D_1\bbE\in H^1(\Real^3_+,\Real^3)$ we remember that
$L(\bbE) = \bbS$ in the weak (distributional) sense which allows us to
express $(D_1)^2\bbE$ in terms of all other second (and lower order)
derivatives of $\bbE$ for which we know that they are in
$L^2(\Real^3_+)$. Therefore, we have $\bbE\in H^2(\Real^3_+,\Real^3)$,
with an estimate
\begin{equation}
\| \bbE \|_{H^2(\Real^3_+)} \leq C_1\left( \| \bbS \|_{L^2(\Real^3_+)}
+ \| \bbg \|_{H^{1/2}(\Real^2)} + \| \bbE \|_{H^1(\Real^3_+)} \right).
\end{equation}

Taking a finite covering of $\Omega$ with a corresponding partition of
unity, and summing up the interior and boundary regularity results, it
follows that $\bbE\in W^2$, and the estimate (\ref{Eq:WkEstimate})
holds for $k=0$. This proves theorem \ref{Thm:Regularity} for $k=0$.

\subsection{Higher order regularity}

Suppose $\bbE\in W^{k+1}$, $k\geq 1$, is a strong solution of
$L(\bbE)_j = S_j$, $b(\bbE)^j = g^j$, where $\bbS\in W^k$, $\bbg\in
H^{k+1/2}(\partial\Omega,\Real^3)$. Suppose also that we have the
estimate
\begin{equation}
\| \bbE \|_{W^{k+1}} \leq C_{k-1}\left( \| L(\bbE) \|_{W^{k-1}}
+ \| b_j(\bbE) \|_{H^{k-1/2}(\partial\Omega)} + \| \bbE \|_{W^k} \right)
\label{Eq:Wk-1Estimate}
\end{equation}
for all $\bbE\in W^{k+1}$. We show that this implies that $\bbE\in
W^{k+2}$. The statement of theorem \ref{Thm:Regularity} then follows
by induction.

The idea of the proof is the same as for $k=0$: We first localize the
problem: Let $\chi$ be the restriction on $\Omega$ of a function in
$C_0^\infty(\Real^3)$. Then $\chi\cdot\bbE\in W^{k+1}$ satisfies the
problem $L(\chi \bbE)_j = \tilde{S}_j$, $b(\chi \bbE)^j = \tilde{g}^j$
where $\tilde{\bbS} \in W^k$ and $\tilde{\bbg}\in
H^{k+1/2}(\partial\Omega,\Real^3)$ are given by Eqs. (\ref{Eq:Stilde})
and (\ref{Eq:gtilde}), respectively. So we can assume that $\bbE$ and
the data are supported in a coordinate patch, and so it is sufficient
to consider the problem on $\Real^3$ or on the half plane $\Real^3_+$.

We discuss only the case $\Real^3_+$ here; interior higher regularity
follows in the same way. We apply the estimate (\ref{Eq:Wk-1Estimate})
to $D_A^h\bbE$, where $A=2,3$ denote tangential directions. Introducing the
commutators
\begin{eqnarray}
R^h(\bbE) &=& L(D_A^h\bbE) - D_A^h L(\bbE), \nonumber\\
r^h(\bbE) &=& b(D_A^h\bbE) - D_A^h b(\bbE), \nonumber
\end{eqnarray}
which satisfy the estimates (which follow from the ``Leibnitz rule''
for difference quotients)
\begin{eqnarray}
\| R^h(\bbE) \|_{H^{k-1}(\Real^3_+)} &\leq& C \| \bbE \|_{H^{k+1}(\Real^3_+)}\; ,
\nonumber\\
\| r^h(\bbE) \|_{H^{k-1/2}(\Real^2)} &\leq& C \| \bbE \|_{H^{k+1/2}(\Real^2)}\; ,
\nonumber
\end{eqnarray}
where the constant $C$ depends on bounds for the metric coefficients $h_{ij}$
and their $k+2$ order derivatives, we obtain
\begin{eqnarray}
\| D_A^h\bbE \|_{H^{k+1}(\Real^3_+)} &\leq& \tilde{C}_k\left( \| D_A^h L(\bbE)  \|_{H^{k-1}(\Real^3_+)}
+ \| D_A^h b(\bbE)_j \|_{H^{k-1/2}(\Real^2)}
+ \| \bbE \|_{H^{k+1}(\Real^3_+)} \right) \nonumber\\
 &\leq& \tilde{C}_k\left( \| \bbS \|_{H^k(\Real^3_+)} + \| \bbg \|_{H^{k+1/2}(\Real^2)} 
+ \| \bbE \|_{H^{k+1}(\Real^3_+)} \right) \nonumber.
\end{eqnarray}
Therefore, $D_A\bbE\in H^{k+1}(\Real^3_+,\Real^3)$, $A=2,3$. Because
$\bbE$ satisfies the equation $L(\bbE) = \bbS$ this also implies
$D_1\bbE\in H^{k+1}(\Real^3_+,\Real^3)$, so $\bbE\in
H^{k+2}(\Real^3_+,\Real^3)$ with a corresponding estimate.



\begin{thebibliography}{10}

\bibitem{FN} 
H. Friedrich and G. Nagy,
Comm. Math. Phys. {\bf 201}, 619 (1999).

\bibitem{CPBC-Stu}
J.M. Stewart,
Class. Quantum Grav. {\bf 15}, 2865 (1998).

\bibitem{CPBC-CLT}
G. Calabrese, L. Lehner, and M. Tiglio, 
Phys. Rev. D {\bf 65}, 104031 (2002).

\bibitem{CPBC-SSW}
B. Szilagyi, B. Schmidt, and J. Winicour,
Phys. Rev. D {\bf 65}, 064015 (2002).

\bibitem{CPBC-SW}
B. Szilagyi and J. Winicour,
Phys. Rev. D {\bf 68}, 041501 (2003).

\bibitem{CPRST}
G. Calabrese, J. Pullin, O. Reula, O. Sarbach and M. Tiglio,
Comm. Math. Phys. {\bf 240}, 377 (2003).

\bibitem{CS}
G. Calabrese and O. Sarbach,
J. Math. Phys. {\bf 44}, 3888 (2003).

\bibitem{CPBC-Frittelli}
S. Frittelli and R. Gomez, Class. Quantum Grav. {\bf 20}, 2379 (2003);
Phys. Rev. D {\bf 68} 044014 (2003); Phys. Rev. D {\bf 69}, 124020 (2004);
gr-qc/0404070.

\bibitem{GMG1}
C. Gundlach and J.M. Mart\'{\i}n-Garc\'{\i}a,
Phys. Rev. D {\bf 70}, 044031 (2004).

\bibitem{GMG2}
C. Gundlach and J.M. Martin-Garc\'{\i}a,
Phys. Rev. D {\bf 70}, 044032 (2004).

\bibitem{LS-FatMax}
L. Lindblom, M.A. Scheel, L.E. Kidder, H.P. Pfeiffer, D. Shoemaker, and S.A. Teukolsky,
Phys. Rev. D {\bf 69}, 124025 (2004).

\bibitem{Kreiss}
H. Kreiss,
Commun. Pure Appl. Math. {\bf 23}, 277 (1970).

\bibitem{KL-Book} 
H.O. Kreiss and J. Lorenz,
{\em ``Initial-Boundary Value Problems and the Navier-Stokes Equations,''} 
Academic Press, (1989).

\bibitem{GKO-Book}
B. Gustafsson, H. Kreiss, and J. Oliger,
{\em ``Time dependent problems and difference methods,''}
John Wiley \& Sons, New York (1995).

\bibitem{AY}
A. Anderson and J.W. York, Jr., 
Phys. Rev. Lett. {\bf 82}, 4384 (1999).

\bibitem{FR}
S. Frittelli and O.A. Reula,
Phys. Rev. Lett. {\bf 76}, 4667 (1996). 

\bibitem{Hern}
S. D. Hern, Ph.D. thesis, University of Cambridge,
1999, gr-qc/0004036.

\bibitem{KST}
L.E. Kidder, M.A. Scheel, and S.A. Teukolsky,
Phys. Rev. D {\bf 64}, 064017 (2001).

\bibitem{Pazy}
See, for instance, A. Pazy, 
{\em "Semigroups of linear operators and applications to partial differential equations",}
Springer-Verlag (1983).

\bibitem{SN}
 M. Shibata and T. Nakamura, Phys. Rev. D {\bf 52}, 5428 (1995).

\bibitem{BS}
T.W. Baumgarte and S.L. Shapiro, Phys. Rev. D {\bf 59}, 024007 (1998)

\bibitem{SCPT}
O. Sarbach, G. Calabrese, J. Pullin, and M. Tiglio, 
Phys. Rev. D {\bf 66}, 064002 (2002).
 
\bibitem{NOR}
G. Nagy, O.E. Ortiz, and O. A. Reula,
Phys. Rev. D {\bf 70}, 044012 (2004).

\bibitem{BeySar}
H. Beyer and O. Sarbach,
{\em On the well posedness of the Baumgarte-Shapiro-Shibata-Nakamura formulation of Einstein's field equations},
gr-qc/0406003.

\bibitem{KWB}
A.M. Knapp, E.J. Walker and T.W. Baumgarte,
Phys. Rev. D {\bf 65}, 064031 (2002).

\bibitem{Fiske}
D.R. Fiske,
Phys. Rev. D {\bf 69}, 047501 (2004).

\bibitem{Calabrese}
G. Calabrese, 
Class. Quantum. Grav. {\bf 21}, 4025 (2004).

\bibitem{BC-preprint}
H. Beyer and M. Chirvasa, {\em Outgoing boundary conditions for evolution systems}, 
(2003), unpublished preprint.

\bibitem{Treves}
F. Treves,
{\em ``Basic linear partial differential equations'',}
Academic Press (1975).

\bibitem{Taylor}
M.E. Taylor, 
{\em ``Partial differential equations, Basic theory'',}
Springer (1999).

\bibitem{RR}
M. Renardy and R.C. Rogers,
{\em ``An Introduction to Partial Differential Equations'',}
Springer-Verlag (1993).

\end{thebibliography}
\end{document}